\documentclass[aps,twocolumn,pra,showpacs,tightenlines,superscriptaddress,10pt]{revtex4-2}
\usepackage[utf8]{inputenc}
\usepackage{amsmath, amssymb, amsfonts}
\usepackage{graphicx}
\usepackage{epsfig}
\usepackage{subfigure}
\usepackage{booktabs}
\usepackage{color}
\usepackage[colorlinks,urlcolor=blue,linkcolor=blue,citecolor=red]{hyperref}
\usepackage[title]{appendix}
\usepackage{etoolbox}
\patchcmd{\appendixname}{Appendix}{APPENDIX}{}{}

\begin{document}
\title{Photon Routing Induced by Giant Atoms in a Synthetic Frequency Dimension}

\author{Ruolin Chai}
\affiliation{Center for Theoretical Physics \& School of Physics and Optoelectronic Engineering, Hainan University, Haikou 570228, China}

\author{Guoqing Cai}
\affiliation{Center for Theoretical Physics \& School of Physics and Optoelectronic Engineering, Hainan University, Haikou 570228, China}

\author{Qiongtao Xie}
\email{xieqiongtao@hainnu.edu.cn}
\affiliation{College of Physics and Electronic Engineering \& Key Laboratory of Laser Technology and Optoelectronic Functional Materials of Hainan Province, Hainan Normal University, Haikou 571158}

\author{Huaizhi Wu}
\email{huaizhi.wu@fzu.edu.cn}
\affiliation{Fujian Key Laboratory of Quantum Information and Quantum Optics \& Department of Physics, Fuzhou University, Fuzhou 350116, China}

\author{Yong Li}
\email{yongli@hainanu.edu.cn}
\affiliation{Center for Theoretical Physics \& School of Physics and Optoelectronic Engineering, Hainan University, Haikou 570228, China}

\date{\today }
\begin{abstract}
We propose a hardware-efficient photon routing scheme based on a dynamically modulated multi-mode ring resonator and a driven cyclic three-level artificial atom, which effectively models a two-level giant atom coupled to a pair of one-dimensional lattices in a synthetic frequency dimension. The routing dynamics of single-photon wave packets in the frequency dimension are investigated numerically and analytically. Our results show that by tuning the phase of the driving field, the photon transmission between the two frequency lattices can be well controlled, thereby determining the propagation direction of photons within the ring resonator. This work presents a feasible scheme for implementing a controllable node in quantum networks, and the predictions of this scheme are well within reach of state-of-the-art experiments. 
\end{abstract}
\maketitle

\section{Introduction}

Quantum routing, which involves the controlled propagation and direction of signals in quantum networks by manipulating their transmission paths, plays an essential role in the field of quantum information~\cite{Quantum.Com,Cirac.1997,kimble2008,zhou2013,lee2022,zhang2022,Cheng2021}. Photons, as ideal carriers of quantum information, offer significant advantages such as high speed, long coherence time, and weak interaction with their environment, making them the preferred choice for flying qubits~\cite{flying2022,flying2024}. Consequently, it has been a crucial and long-term goal to explore efficient photon routing mechanisms in quantum information. This is typically achieved by controlling photon absorption and transmission with quantum emitters.

Recently, a novel class of quantum emitters, known as giant atoms~\cite{Giant.atom.summarize}, has been proposed and widely studied. Unlike natural atoms that are coupled to propagating fields (e.g. electromagnetic field) at a single point, giant atoms are able to interact with the fields at multiple separate points. This feature not only provides a new paradigm for light-atom interactions~\cite{SAW.Gustafsson,meandering.Kockum}, but also has potential applications in, e.g., photon routing~\cite{zhang2022,Cheng2021}, quantum simulation of many-body physics~\cite{chenGZ2024}, and quantum information processing~\cite{NM.Xu2024}. Giant atom systems can be implemented on various platforms by coupling artificial atoms to propagating fields, such as surface acoustic waves~\cite{SAW.Andersson2019,SAW.Andersson2020,SAW.Guo,SAW.Manenti,SAW.Noguchi,SAW.Qiu}, microwaves in meandering waveguides~\cite{meandering.Kannan,meandering.Vadiraj}, and fields propagating in optical lattices~\cite{Du_2023,Wang2021}, at multiple coupling points. The phase accumulated among different coupling points leads to unique self-interference phenomena that are not typically observed in natural atoms, such as frequency-dependent Lamb shifts~\cite{meandering.Kockum}, in-band decoherence-free interactions~\cite{DFI.Carollo,DFI.Kockum,DFI.Lei,DFI.Soro,DFI.soro.2023}, non-Markovian dynamics~\cite{NM.andersson,NM.Du,NM.Du2023,NM.Yin,DFI.soro.2023} and chiral spontaneous emission and scattering~\cite{Chiral.Wang,chiralWu2024,Chen2024,Zhao2020,chen2022,Xu2025,weng2024,zhu2024,wang2024}. 

Synthetic dimensions refer to artificial coordinates that are simulated using non-spatial degrees of freedom, enabling the study of high-dimensional phenomena in a physically simple system~\cite{SD.Yuan}. Common platforms for realizing synthetic dimensions include cold atoms~\cite{ColdAtom.Celi,ColdAtom.Price,ColdAtom.Price2015}, superconducting quantum circuits~\cite{SQC.Feng,SQC.Tsomokos}, and photonic systems~\cite{PS.Zhou,PS.Ozawa,PS.Armandas}. These candidates hold promise for engineering and manipulating light-matter interactions in a hardware-efficient and programmable manner. In photonic systems, for example, synthetic dimensions are often realized through various mechanisms, such as resonator frequency modes~\cite{FD.Yuan,FD.Qin,FD.Qian,FD.Yuan18,FD.Yuan2018,FD.yuan2021}, the orbital angular momenta of photon~\cite{OAM.Yuan,PS.lustig,PS.luo,OAM.Mu}, and the time sequences of multi-pulse~\cite{MP.Yuan,MP_Alois,MP_wimmer2013,MP_wimmer2015}. Notably, by coupling the energy-level transitions of an artificial atom with different frequency modes in a synthetic frequency dimension, a ``nonlocal'' interaction between the atom and the frequency lattice (FL) can be realized~\cite{GiantAtom.Lei}. This approach provides an effective framework for investigating phenomena analogous to those observed in giant atoms.

In this work, we propose a scheme for photon routing that leverages the unique properties of giant atoms and synthetic frequency dimensions to efficiently scatter flying qubits to specific lattice. By interacting a driven artificial atom~\cite{SQC.Murali,SQC.Wallraff} with a dynamically modulated multi-mode ring resonator~\cite{Broadband2019,frequencycomb}, we can effectively model a giant atom coupled to a pair of 1D FLs in the frequency dimension. We numerically and analytically study the single-photon scattering behavior in this synthetic giant-atom system, demonstrating efficient and targeted photon routing in the frequency dimension. 

\section{System and Model}

\begin{figure}[t]
\centering \subfigure{ \includegraphics[width=0.95\linewidth]{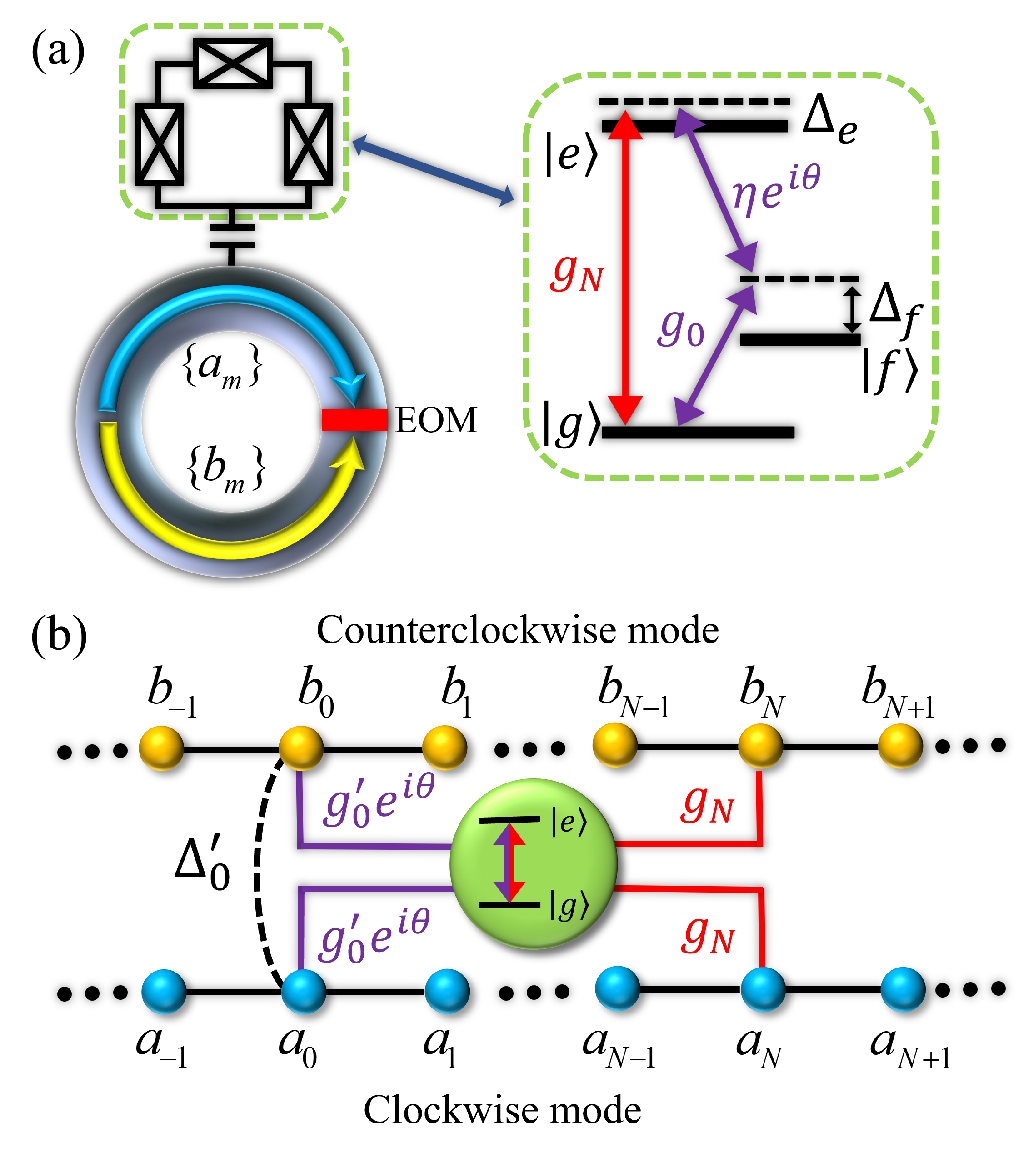}\label{fig:1(a)}}
\subfigure{\label{fig:1(b)}}\caption{(a) Schematic diagram for a superconducting quantum circuit coupled to the ring resonator modulated by an electro-optic modulator (EOM). A cyclic $\Delta$-type three-level artificial atom interacts with two kinds of different-frequency photon modes of the ring resonator (with the coupling strengths $g_{0}$ and $g_{N}$) and is driven by an external field (with the coupling strength $\eta$). (b) Effective giant-atom model in the synthetic frequency dimension. $a_{m}\ (b_{m})$ represents the $m$-th clockwise (counterclockwise) photon frequency mode in the ring resonator or the $m$-th site of $a$-FL ($b$-FL). $g_{0}^{\prime}=g_{0}\eta/\Delta_{f}$ is the effective coupling strength between the lattice site $a_{0}$ (or $b_{0}$) and transition $\left|g\right\rangle \leftrightarrow\left|e\right\rangle $. $\Delta_{0}^{\prime}=g_{0}^{2}/\Delta_{f}$ is the induced coupling strength between them arising from the adiabatic elimination.}
\end{figure}

First, let us consider a multi-mode ring resonator modulated by an electro-optic modulator (EOM), as illustrated in Fig.~\ref{fig:1(a)}. The ring resonator has a series of modes that are equally spaced in frequency within a specific frequency range. By setting the modulation frequency to match the free spectral range $\Omega_{s}$ of the resonator, the EOM introduces coherent coupling between adjacent resonant frequency modes of the resonator~\cite{PS.Armandas,xiao2022bound,Broadband2019,frequencycomb}. As a result, the modulated ring resonator supports a pair of 1D FLs, denoted by $a$-FL and $b$-FL hereafter, which correspond to two similar sets of whispering-gallery modes propagating in clockwise and counterclockwise directions, respectively. In the interaction picture, the Hamiltonian of the 1D FLs can be written as ($\hbar=1$)~\cite{FD.yuan2021,FD.Yuan,FD.Yuan18,FD.Yuan2018,FD.Qian,FD.Qin}

\begin{equation}
H_{\text{FL}}=-J\sum_{m}\left(\hat{a}_{m}^{\dagger}\hat{a}_{m+1}+\hat{b}_{m}^{\dagger}\hat{b}_{m+1}+\text{H.c.}\right),\label{eq:1}
\end{equation}
where $\hat{a}_{m}\ (\hat{a}_{m}^{\dagger})$ and $\hat{b}_{m}\ (\hat{b}_{m}^{\dagger})$ are the annihilation (creation) operators for the $m$-th modes in the clockwise and counterclockwise directions (i.e. the $m$-th sites of the two FLs), respectively, with frequency $\omega_{m}=\omega_{0}+m\Omega_{s}$ $(m=0,\pm1,\pm2,...)$. Here $\omega_{0}$ is the frequency of the central (clockwise and counter-clockwise) modes within the considered frequency range. Hamiltonian (\ref{eq:1}) describes the hopping between adjacent modes $a_{m}\ (b_{m})$ and $a_{m+1}\ (b_{m+1})$ modulated by the EOM, with a real coupling strength $J$. This coupling strength can be controlled by adjusting the modulation amplitude of the EOM.

Next, an additional superconducting quantum device comprising a loop with three Josephson junctions~\cite{SQC.Wallraff} is coupled to the modulated ring resonator via a capacitor. This device functions as a cyclic three-level artificial atom, consisting of a ground state $\left|g\right\rangle $, an excited state $\left|e\right\rangle $, and an intermediate state $\left|f\right\rangle $. By taking the energy of the ground state level $\omega_{g}=0$ as a reference, the energy of the intermediate (excited) state is denoted by $\omega_{f}$ ($\omega_{e}$). The transitions $\left|g\right\rangle \leftrightarrow\left|f\right\rangle$ and $\left|g\right\rangle\leftrightarrow\left|e\right\rangle$ are coupled to the frequency modes $a_{0}\ (b_{0})$ and $a_{N}\ (b_{N})$, with coupling strengths $g_{0}$ and $g_{N}$, respectively. $N$ is the coupling separation between the two points at which the giant atom couples with each FL. Moreover, the transition $\left|f\right\rangle \leftrightarrow\left|e\right\rangle$ is driven by an external field with frequency $\omega_{d}$, amplitude $\eta$, and a driving phase $\theta$. Under the three-photon resonance condition $\omega_{d}+\omega_{0}=\omega_{N}$, the system Hamiltonian can be given in a time-independent form, i.e., 
\begin{eqnarray}
H & = & H_{\text{FL}}-\Delta_{e}\left|e\right\rangle \left\langle e\right|-\Delta_{f}\left|f\right\rangle \left\langle f\right|\nonumber \\
 &  & +\left[g_{N}\left|g\right\rangle \left\langle e\right|\left(\hat{a}_{N}^{\dagger}+\hat{b}_{N}^{\dagger}\right)+g_{0}\left|g\right\rangle \left\langle f\right|\left(\hat{a}_{0}^{\dagger}+\hat{b}_{0}^{\dagger}\right)\right.\nonumber \\
 &  & \left.\hspace{0.8em}+\eta e^{i\theta}\left|e\right\rangle \left\langle f\right|+\text{H.c.}\right]\text{.}\label{eq:2}
\end{eqnarray}
Here $\Delta_{e}=\omega_{e}-\omega_{N}$ is the detuning between the transition $\left|g\right\rangle \leftrightarrow\left|e\right\rangle$ and the frequency mode $a_{N}\ (b_{N})$. Similarly, $\Delta_{f}=\omega_{f}-\omega_{0}$ represents the detuning between the transition $\left|g\right\rangle \leftrightarrow\left|f\right\rangle$ and the frequency mode $a_{0}\ (b_{0})$. The coupling strengths ($g_{0}$, $g_{N}$, and $\eta$) are assumed to be real without loss of generality, while the driving phase $\theta$, determined by the external field, plays a crucial role in controlling the photon scattering and routing behavior, as will be shown below. In Sec.~\ref{subsec:3A}, we will elucidate that Hamiltonian~\eqref{eq:2} can effectively model a two-level giant atom coupled to each of the two FLs at two different sites in the frequency dimension. 

For simplicity, here we focus on the system in the single-excitation subspace, which is described by the state
\begin{equation}
\left|\psi(t)\right\rangle =\sum_{m}\left[u_{m}(t)\hat{a}_{m}^{\dagger}+v_{m}(t)\hat{b}_{m}^{\dagger}\right]\left|0,g\right\rangle +\sum_{\beta=e,f}w_{\beta}(t)\left|0,\beta\right\rangle .\label{eq:3}
\end{equation}
Here, $\left|0,\alpha\right\rangle $ ($\alpha=g,e,f$) represents the state where the FLs remain in the vacuum and the artificial atom is in state $\left|\alpha\right\rangle $. $u_{m}\left(t\right)$ and $v_{m}\left(t\right)$ represent the probability amplitudes for a photon being in the $m$-th site of $a$-FL and $b$-FL, respectively. Similarly, $w_{\beta}\left(t\right)$ represents the excitation probability amplitudes of the giant atom in the state $\left|\beta\right\rangle $.

\section{Photon Routing in the Frequency Dimension\protect\label{sec:III}}

Recently, it has been shown that by interacting a superconductive transmission line with an artificial atom with a driving field, one can mimic the model of a giant atom coupling to a single FL in synthetic frequency dimension~\cite{GiantAtom.Lei}. The phase of the driving field can introduce a nontrivial coupling phase difference between the two coupling points of the synthetic giant-atom system, thus enabling the control of scattering behavior in the giant atom. This approach offers a significant advantage over real-space correspondence, which typically require complex time-dependent modulation of the magnetic flux to control the coupling phase~\cite{gong2024}. In contrast, the synthetic model allows direct phase manipulation through the driving field.  Building on this insight, we aim to develop a synthetic giant atom model where the giant atom is coupled to a pair of 1D FLs. By tuning the driving phase, this setup allows on-demand control over the scattering behavior of incident photons, enabling targeted routing between the two FLs.

\subsection{Numerical Simulation\protect\label{subsec:3A}}

To investigate the single-photon routing properties of our system, we numerically solve the Schrödinger equation, which gives rise to 
\begin{eqnarray}
i\dot{w}_{e} & = & \left(-i\gamma_{e}-\Delta_{e}\right)w_{e}+g_{N}\left(u_{m}+v_{m}\right)\delta_{m,N}+\eta e^{i\theta}u_{f},\nonumber \\
i\dot{w}_{f} & = & \left(-i\gamma_{f}-\Delta_{f}\right)w_{f}+g_{0}\left(u_{m}+v_{m}\right)\delta_{m,0}+\eta e^{-i\theta}w_{e},\nonumber \\
i\dot{u}_{m} & = & -J\left(u_{m+1}+u_{m-1}\right)+g_{N}w_{e}\delta_{m,N}+g_{0}w_{f}\delta_{m,0},\nonumber \\
i\dot{v}_{m} & = & -J\left(v_{m+1}+v_{m-1}\right)+g_{N}w_{e}\delta_{m,N}+g_{0}w_{f}\delta_{m,0}.\label{eq:4}
\end{eqnarray}
Here, the terms $-i\gamma_{e}w_{e}$ and $-i\gamma_{f}w_{f}$ are introduced phenomenologically, where $\gamma_{e}$ and $\gamma_{f}$ represent the intrinsic dissipation rates of states $\left|e\right\rangle $ and $\left|f\right\rangle $, respectively. For a typical superconducting circuit setup, the intrinsic dissipation rates $\gamma_{e}$ and $\gamma_{f}$ can be much smaller than the other parameters~\cite{meandering.Kannan} (e.g. $g_{0},\ g_{N},\ \eta,\ J$) and are thus neglected in the subsequent analysis (i.e., $\{\gamma_{e},\gamma_{f}\}\rightarrow0$).

Assume that the detuning $\Delta_{f}$ is much larger than the single-photon coupling strengths and the transition $\left|g\right\rangle \leftrightarrow\left|e\right\rangle$ is nearly resonant with the $N$-th modes, i.e., $\{\eta,\ g_{0}\}\ll\Delta_{f}$ and $\Delta_{e}\ll g_{N}$, the intermediate state $\left|f\right\rangle$ can be adiabatically eliminated, provided it is initially unpopulated. Under these conditions, Eq.~\eqref{eq:4} reduces to 
\begin{eqnarray}
i\dot{w}_{e} & = & \left(\Delta_{e}^{\prime}-\Delta_{e}-i\gamma_{e}\right)w_{e}+g_{0}^{\prime}e^{i\theta}\left(u_{0}+v_{0}\right)\nonumber \\
 &  & +g_{N}\left(u_{N}+v_{N}\right),\nonumber \\
i\dot{u}_{m} & = & -J\left(u_{m+1}+u_{m-1}\right)+g_{N}w_{e}\delta_{m,N}\nonumber \\
 &  & +\left(g_{0}^{\prime}e^{-i\theta}w_{e}+\Delta_{0}^{\prime}u_{0}+\Delta_{0}^{\prime}v_{0}\right)\delta_{m,0},\nonumber \\
i\dot{v}_{m} & = & -J\left(v_{m+1}+v_{m-1}\right)+g_{N}w_{e}\delta_{m,N}\nonumber \\
 &  & +\left(g_{0}^{\prime}e^{-i\theta}w_{e}+\Delta_{0}^{\prime}u_{0}+\Delta_{0}^{\prime}v_{0}\right)\delta_{m,0}.\label{eq:5}
\end{eqnarray}
Here $\Delta_{e}^{\prime}=\eta^{2}/\Delta_{f}$ is the effective frequency (Lamb) shift of the excited state $\left|e\right\rangle $. $\Delta_{0}^{\prime}=g_{0}^{2}/\Delta_{f}$ is the effective frequency shift of the lattice sites $a_{0}$ and $b_{0}$, as well as the induced coupling strength between them arising from the adiabatic elimination (the black dashed line in Fig.~\ref{fig:1(b)}). $g_{0}^{\prime}=g_{0}\eta/\Delta_{f}$ is the effective coupling strength between the lattice site $a_{0}$ (or $b_{0}$) and transition $\left|g\right\rangle \leftrightarrow\left|e\right\rangle$. The driving phase $\theta$ is integrated into the effective coupling phase during the adiabatic elimination process. The equations of motion in Eq.~\eqref{eq:5} describe a two-level ``giant atom'' coupled to a pair of 1D FLs at separated sites, as depicted in Fig.~\ref{fig:1(b)}. In what follows, we set $\Delta_{e}^{\prime}=\Delta_{e}$ to effectively offset the detuning between transition $\left|g\right\rangle \leftrightarrow\left|e\right\rangle$ and lattice sites $a_{N}$ (or $b_{N}$). In a standard giant-atom model where a giant atom interacts with a waveguide at two coupling points, the two coupling strengths are usually assumed to be equal in order to realize ideal interference effects. In view of this, we always set $g_{0}^{\prime}=g_{N}=g$ (i.e. $g_{0}\eta/\Delta_{f}=g_{N}=g$) in the following discussions, which can be achieved by carefully tuning the parameters in practical experiments.

To study the scattering behavior of our model, we consider a single-photon wave packet coming from the left side of $a$-FL, with the Gaussian type 
\begin{equation}
\left|\psi(0)\right\rangle =A\sum_{m}\exp\left[-\frac{\left(m-m_{0}\right)^{2}}{2\sigma^{2}}+ik_{\mathrm{f}}m\right]\hat{a}_{m}^{\dagger}\left|0,g\right\rangle ,\label{eq:6}
\end{equation}
where $A$ is the normalization factor and $\sigma$ represents the width of the Gaussian wave packet in the frequency dimension. The ``wave vector'' in frequency dimension $k_{\mathrm{f}}>0$ corresponds to the ``rightward'' propagation of the wave packet along the FL. The wave packet is centered at ``$m_{0}<0$ and $\left|m_{0}\right|\gg1$'', indicating that it is incident far from the left of the coupling sites in $a$-FL (throughout this work, we always consider such an incident case). In this section, we assume that the width of the wave packet is much larger than the coupling separation between the coupling points ($\sigma\gg N$) to avoid the non-Markovian retardation effects. The sum of the excitation probabilities over $b$-FL ($a$-FL) in the long-time limit is defined as the transmission coefficient at $b$-FL ($a$-FL), i.e., $T_{b}=\sum_{m=N}^{\infty}\left|v_{m}(t=+\infty)\right|^{2}(T_{a}=\sum_{m=N}^{\infty}\left|u_{m}(t=+\infty)\right|^{2})$.

\begin{figure}[htbp]
\centering \subfigure{\includegraphics[width=0.85\linewidth]{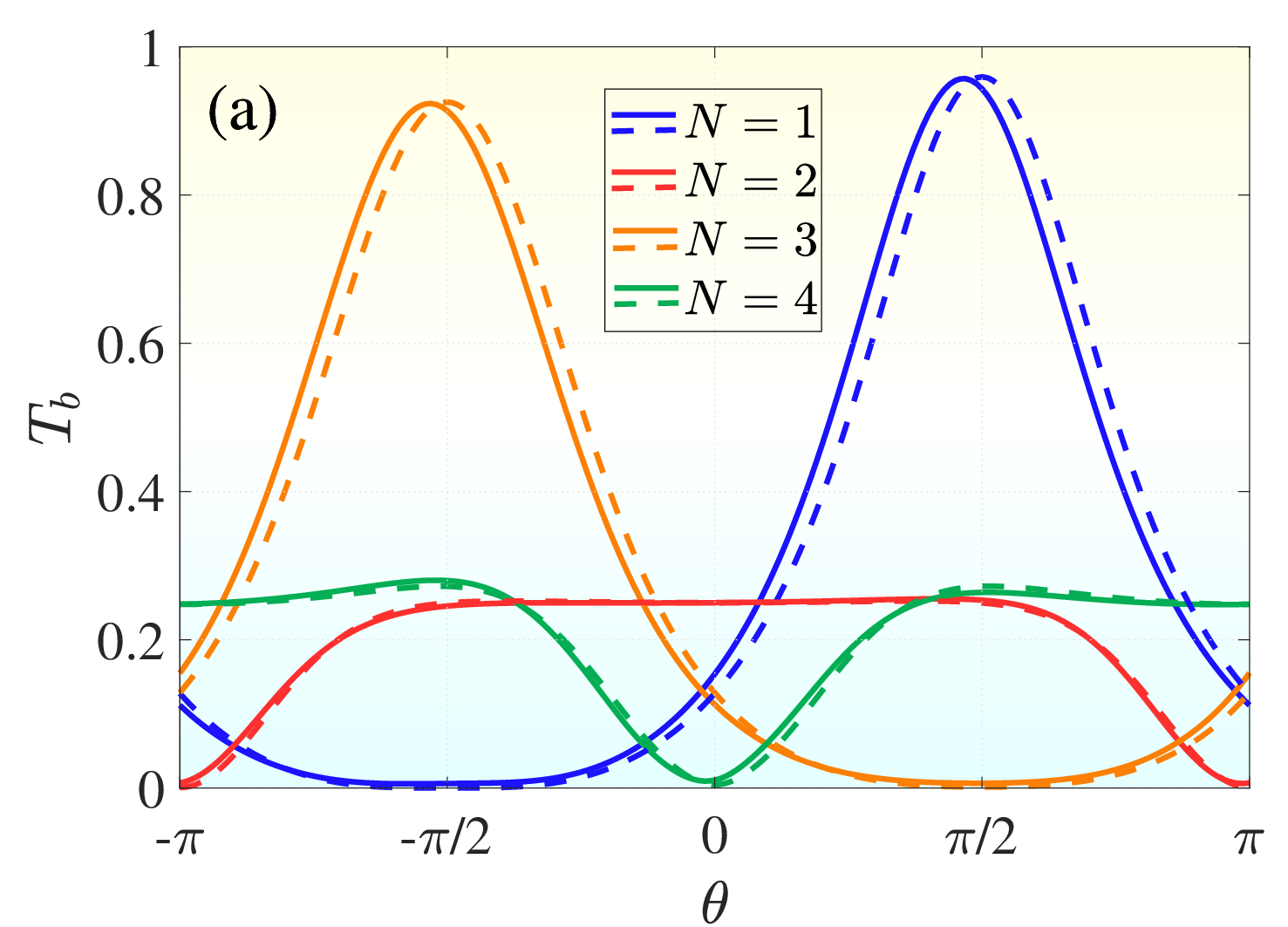}\label{fig:2(a)}}
\subfigure{\includegraphics[width=0.85\linewidth]{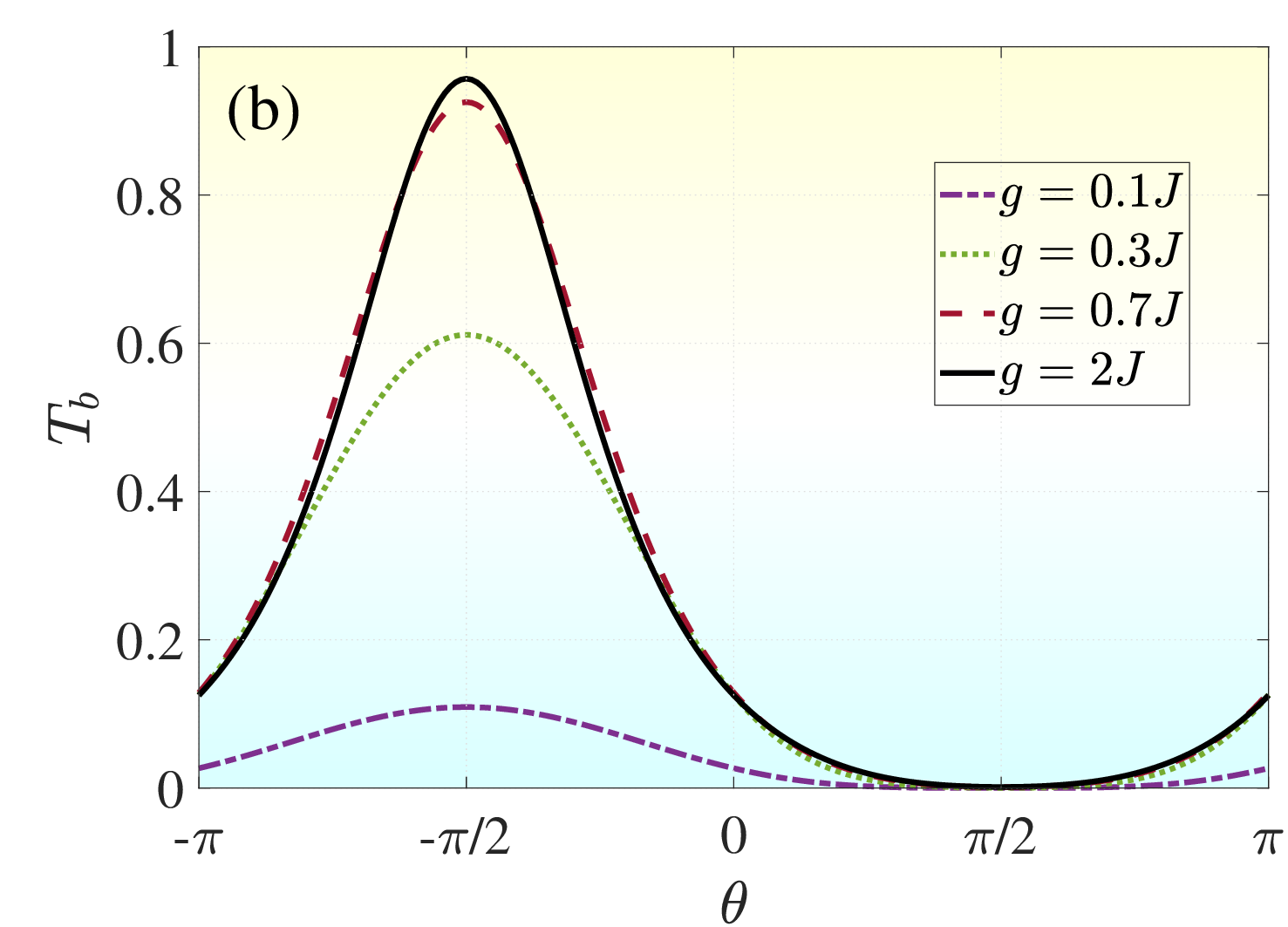}\label{fig:2(b)}}
\caption{(a) The transmission coefficient for photon routing at $b$-FL. The solid and dashed lines represent the cases with and without considering the $\Delta_{0}^{\prime}$ terms in Eq.~\eqref{eq:5}, respectively. (b) The transmission coefficient on $b$-FL versus the driving phase $\theta$ for different coupling strengths. In panel (a), the parameters are $g_{0}=4J$, $g_{N}=0.7J$, $\eta=17.5J$, and $\Delta_{f}=100J$. In panel (b), the parameters are set as $N=3$, $g_{0}^{\prime}=g_{N}=g$, and the $\Delta_{0}^{\prime}$ terms in Eq.~\eqref{eq:5} are neglected. The other parameters are $\sigma=20$ and $k_{\mathrm{f}}=\pi/2$.}
\end{figure}

\begin{figure*}[htbp]
\centering \subfigure{\includegraphics[width=0.32\linewidth]{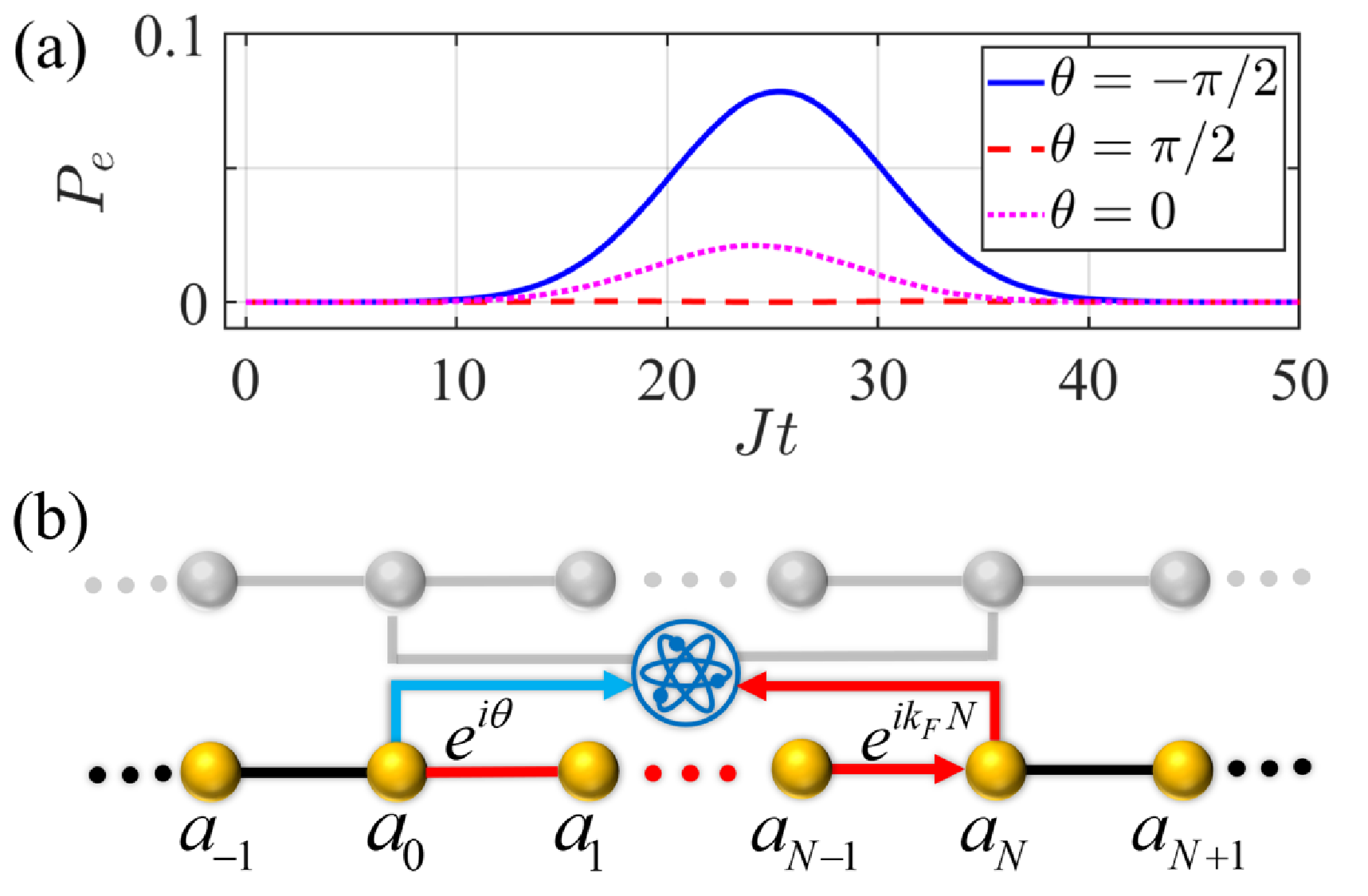}\label{fig:3(a)}}
\subfigure{\label{fig:3(b)}} \subfigure{\includegraphics[width=0.32\linewidth]{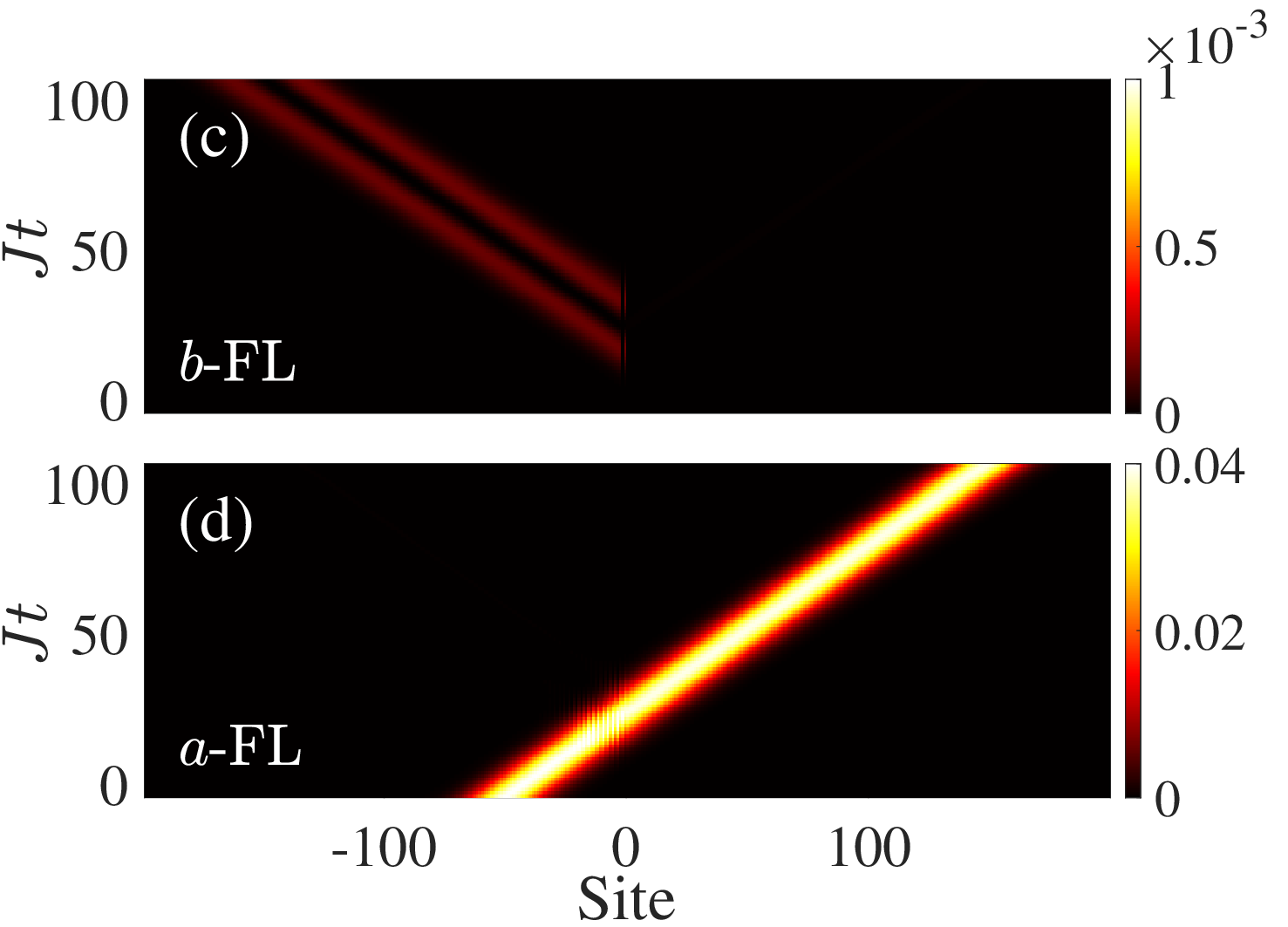}\label{fig:3(c)}}
\subfigure{\label{fig:3(d)}} \subfigure{\includegraphics[width=0.32\linewidth]{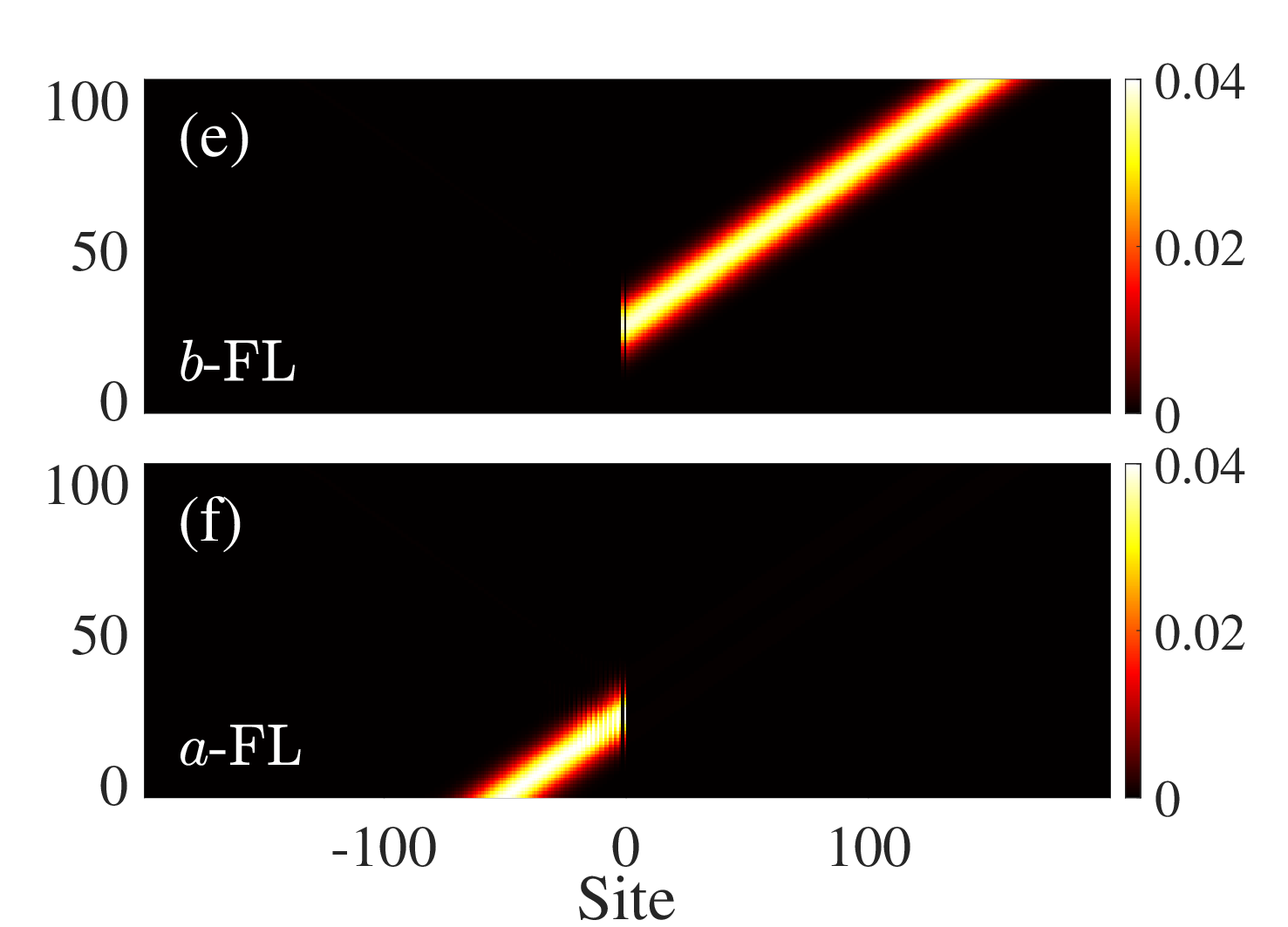}\label{fig:3(e)}}
\subfigure{\label{fig:3(f)}} \caption{(a) Time evolution of the excitation probability of the giant atom for different driving phase. (b) Two excitation paths of the giant atom in $a$-FL. The phase of the blue path is contributed by the driven field, while the phase of the red path is contributed by the accumulated phase from the coupling separations N. Here, we only analyze the excitation process in $a$-FL.  (c)-(f) Time evolution of the field probability distributions in $a$-FL and $b$-FL. In panels (c) and (d), the driving phase is set to be $\theta=\pi/2$ . In panels (e) and (f), the driving phase is set to be $\theta=-\pi/2$. Other parameters are $N=3$, $g_{0}=4J,g_{N}=0.7J$, $\eta=17.5J$, $\Delta_{f}=100J$, $\sigma=20$, and $k_{\mathrm{f}}=\pi/2$.}
\end{figure*}

For the numerical simulations, we select 400 equally spaced frequency modes in a specific microwave frequency range~\cite{PS.Tzuang,Broadband2019} (i.e., 400 sites in total for each FL) and assume that the giant atom is located roughly at the middle of the FLs with a small coupling separation $N$ to avoid the boundary effect. Using Eq.~\eqref{eq:5}, we show the numerical results of $T_{b}$ as a function of the driving phase $\theta$ for different coupling separation $N$ in Fig.~\ref{fig:2(a)}. Here the parameters are set as $g_{0}=4J$, $\eta=17.5J$, $\Delta_{f}=100J$, and $g_{N}=0.7J$, which implies $\Delta_{0}^{\prime}=0.16J$ and $g_{0}^{\prime}=g_{N}=0.7J$. Specifically, for odd values of $N$, such as $N=1$, the transmission coefficient $T_{b}\approx0.96$ when $\theta=\pi/2$. In contrast, for even values of $N$, the transmission coefficient $T_b$ is also dependent on $\theta$, but its maximal value becomes smaller. For example, when $N=2,4$, the transmission coefficient $T_{b}$ is less than approximately $0.25$. The results show that the transmission coefficient of the Gaussian wave packet can be effectively controlled by the driving phase $\theta$, particularly when $N$ is odd. 

In Fig.~\ref{fig:2(a)}, we also investigate the impact of the frequency shift $\Delta_{0}^{\prime}$ on the transmission coefficient. By comparing the transmission coefficient curves with (solid lines) and without (dashed lines) the terms containing $\Delta_{0}^{\prime}$ in Eq.~\eqref{eq:5}, we find a slight deviation when these terms containing $\Delta_{0}^{\prime}$ are included. Such a deviation, which mainly arises from the additional coupling between sites $a_{0}$ and $b_{0}$, suggests that neglecting the detuning terms in Eq.~\eqref{eq:5} is a justified approximation. Therefore, we disregard the terms containing $\Delta_{0}^{\prime}$ in the following discussions to simplify the analysis.

Fig.~\ref{fig:2(b)} shows the transmission coefficient $T_{b}$ as a function of the driving phase $\theta$, with different coupling strengths between the giant atom and FLs. Here, we consider the coupling separation $N=3$ as an example. When the driving phase $\theta=-\pi/2$, the photon routing probability is significantly enhanced as the coupling strength increases, especially when  $g<0.5J$. Interestingly, for $\theta=\pi/2$, the transmission coefficient $T_{b}$ remains zero regardless of the value of the effective coupling strength $g$. 

Now we would like to analyze the interaction process between the incident wave packet and the giant atom. Figure~\ref{fig:3(a)} illustrates the time evolution of the atomic excitation probability $P_{e}=|w_{e}|^{2}$ of state $\left|e\right\rangle $, for different driving phases $\theta$ and with the coupling separation $N=3$. For $k_{\mathrm{f}}=\pi/2$, the phase accumulated between the two coupling points is $k_{\mathrm{f}}N=3\pi/2$, determined by the ``wave vector'' of the wave packet. For the case of $\theta=\pi/2$, the phase difference between the two excitation paths of the giant atom [through the two coupling points, as shown by Fig.~\ref{fig:3(b)}] is $k_{\mathrm{f}}N-\theta=\pi$, leading to a destructive interference so that the giant atom is barely excited [see the red line in Fig.~\ref{fig:3(a)}]. As shown in Fig.~\ref{fig:3(c)}, only a small fraction of the wave packet is scattered into the $b$-FL due to the weak non-Markovian retardation effect~\cite{NM.Xu2024}. The non-Markovian retardation effect results from the finite time required for the wave packet to propagate between the two coupling points. Since $N$ is relatively small, the excitation scattered into the $b$-FL can be neglected. In contrast, for the case of $\theta=-\pi/2$, the phase difference between the two paths becomes $2\pi$, leading to a constructive interference [see the blue line in Fig.~\ref{fig:3(a)}]. That is, in this case, the giant atom can be efficiently excited by the incident wave packet, and then emit photons towards $b$-FL.

To gain a clearer understanding of the photon scattering dynamics induced by the giant atom, we examine the probability distribution across the entire FLs. As depicted in Figs.~\ref{fig:3(c)} and \ref{fig:3(d)}, when $\theta=\pi/2$, the majority of wave packet continues to propagate along the $a$-FL, without significant effect from the giant atom due to destructive interference. However, as shown in Figs.~\ref{fig:3(e)} and \ref{fig:3(f)} where $\theta=-\pi/2$, most of the excitations are routed to $b$-FL. Based on the above analysis, it is clear that the wave packet incident from $a$-FL can be selectively scattered to different FLs depending on the driving phase of the external field, thereby achieving controllable photon routing in the frequency dimension.

\subsection{Analytical Solution}

To better elucidate the scattering behavior and the adaptability of our model, we analytically study the effective model described by Eq.~\eqref{eq:5} with single-photon ``plane waves'' incident from the left side of $a$-FL. For simplicity, we again set $g_{0}^{\prime}=g_{N}=g$ and neglect the terms containing $\Delta_{0}^{\prime}$ (whose effect has been shown to be negligible in Sec.~\ref{subsec:3A}). According to the stationary Schrödinger equation $H\left|\psi\right\rangle =E_{\mathrm{f}}\left|\psi\right\rangle$ in frequency dimension, where the stationary state in the single-excitation subspace is expressed as $\left|\psi\right\rangle =\sum_{m}\left(u_{m}\hat{a}_{m}^{\dagger}+v_{m}\hat{b}_{m}^{\dagger}\right)\left|0,g\right\rangle +w_{e}\left|0,e\right\rangle$, we obtain
\begin{eqnarray}
E_{\mathrm{f}}w_{e} & = & e^{i\theta}g\left(u_{0}+v_{0}\right)+g\left(u_{N}+v_{N}\right),\nonumber \\
E_{\mathrm{f}}u_{m} & = & -J\left(u_{m+1}+u_{m-1}\right)+e^{-i\theta}gw_{e}\delta_{m,0}+gw_{e}\delta_{m,N},\nonumber \\
E_{\mathrm{f}}v_{m} & = & -J\left(v_{m+1}+v_{m-1}\right)+e^{-i\theta}gw_{e}\delta_{m,0}\nonumber \\
 &  & +gw_{e}\delta_{m,N}.\label{eq:7}
\end{eqnarray}
By using the Bethe ansatz~\cite{levkovich2016bethe}, the probability amplitudes $u_{m}$ and $v_{m}$ can be written as 
\begin{eqnarray}
u_{m} & = & \begin{cases}
e^{ik_{\mathrm{fa}}m}+r_{a}e^{-ik_{\mathrm{fa}}m}, & m<0\\
l_{al}e^{-ik_{\mathrm{fa}}m}+l_{ar}e^{ik_{\mathrm{fa}}m}, & 0\leqslant m\leqslant N\\
t_{a}e^{ik_{\mathrm{fa}}m}, & m>N
\end{cases}\label{eq:8}
\end{eqnarray}
and 
\begin{eqnarray}
v_{m} & = & \begin{cases}
r_{b}e^{-ik_{\mathrm{fb}}m}, & m<0\\
l_{bl}e^{-ik_{\mathrm{fb}}m}+l_{br}e^{ik_{\mathrm{fb}}m}, & 0\leqslant m\leqslant N\\
t_{b}e^{ik_{\mathrm{fb}}m}, & m>N
\end{cases}\label{eq:9}
\end{eqnarray}
where $r_{a}\left(r_{b}\right)$ and $t_{a}\left(t_{b}\right)$ are, respectively, the single-photon reflection and transmission amplitudes by the giant atom in the $a$-FL ($b$-FL). $l_{al}$ ($l_{bl}$) and $l_{ar}$ ($l_{br}$) represent the probability amplitudes for photons propagating leftward and rightward between the two coupling sites of $a$-FL ($b$-FL), respectively. We consider the real-space correspondence of Eq.~\eqref{eq:5}, the photon absorbed by the atom should have the same wave vector as the emitted one. Therefore, in the frequency dimension, we also assume that photons propagating in both FLs share the same wave vector,i.e., $k_{\mathrm{fa}}=k_{\mathrm{fb}}=k_{\mathrm{f}}$. This assumption is analogous to the scenario in real-space lattices, where the photons absorbed by the giant atom should have the same energy as those emitted. By applying the continuity conditions at $m=0$ and $m=N$, we obtain the amplitudes (see Appendix~\ref{APPB} for detailed calculations) 
\begin{equation}
t_{a}=\frac{[1+\cos(k_{\mathrm{f}}N+\theta)]+i\xi}{[\cos (k_{\mathrm{f}}N)\cos\theta+1]+i\xi},\label{eq:10}
\end{equation}
\begin{equation}
t_{b}=-\frac{[1+\cos(k_{\mathrm{f}}N-\theta)]}{2[\cos (k_{\mathrm{f}}N)\cos\theta+1]+i\xi},\label{eq:11}
\end{equation}
where $\xi=2\sin (k_{\mathrm{f}}N)\cos\theta+J^{2}\sin(2k_{\mathrm{f}})/g^{2}$. To distinguish from the transmission coefficients obtained numerically, here we define $\tilde{T}_{a}=\left|t_{a}\right|^{2}\ (\tilde{R}_{a}=\left|r_{a}\right|^{2})$ and $\tilde{T}_{b}=\left|t_{b}\right|^{2}$ ($\tilde{R}_{b}=\left|r_{b}\right|^{2}$) as the transmission (reflection) coefficients of the incident plane wave propagating in the $a$-FL and $b$-FL, respectively.

\begin{figure}
\centering \subfigure{\includegraphics[width=0.85\linewidth]{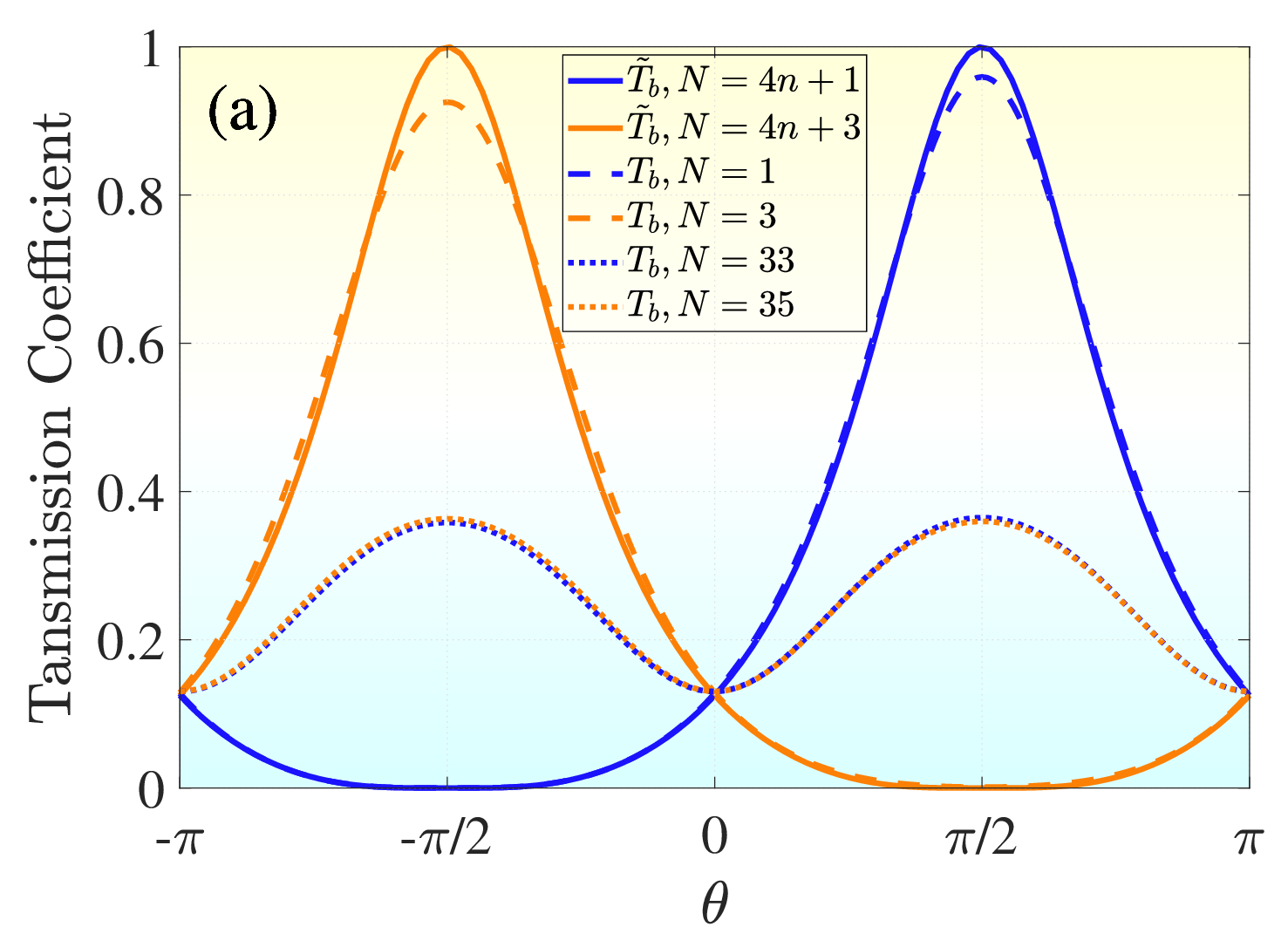}\label{fig:4(a)}}
\subfigure{\includegraphics[width=0.85\linewidth]{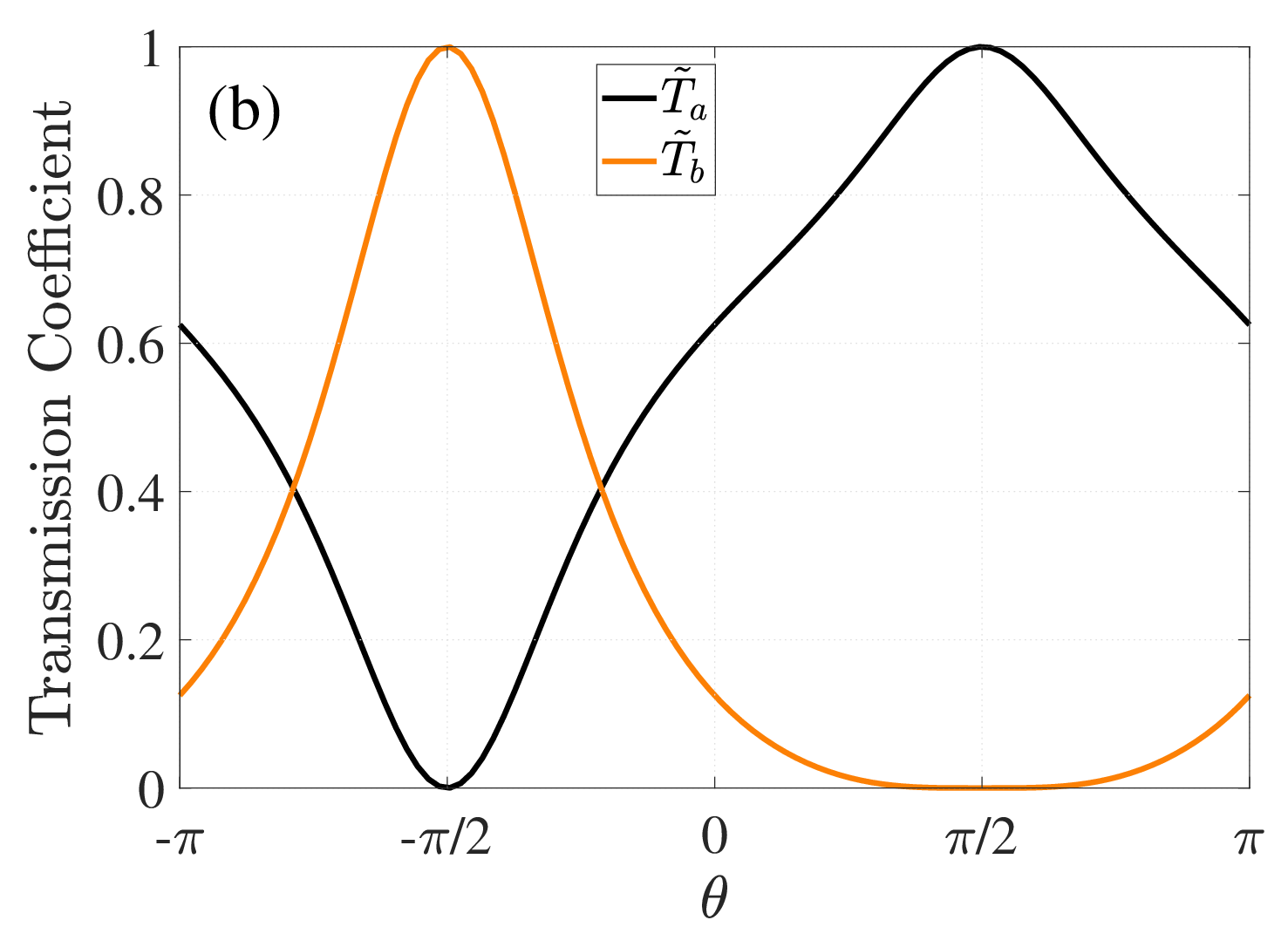}\label{fig:4(b)}}
\caption{(a) The transmission coefficients $T_{b}$ and $\tilde{T}_{b}$ as a function of the driving phase $\theta$. The solid lines represent analytical solutions of Eq.~\eqref{eq:11}. The dashed lines represent numerical solutions of Eq.~\eqref{eq:5} with a Gaussian wave packet excitation, and the dotted lines indicate the numerical solutions where the Gaussian wave packet width is much smaller than the spacing between coupling separation. (b) The transmission coefficients $\tilde{T_{a}}$ and $\tilde{T}_{b}$ versus the driving phase $\theta$ for $N=4n+3$ $(n\in\mathbb{N})$. In panel (a) the parameters are set as $g=0.7J$, $k_{\mathrm{f}}=\pi/2$, $\eta=17.5J$,
$\Delta_{f}=100J$, $\sigma=20$ for numerical solutions, $g=0.7J$,
$k_{\mathrm{f}}=\pi/2$ for analytical solutions. In panel (b) the parameters
are set as $g=0.7J$ and $k_{\mathrm{f}}=\pi/2$.}
\end{figure}

We plot in Fig.~\ref{fig:4(a)} both the numerical and analytical transmission coefficients $T_{b}$ and $\tilde{T}_{b}$ as a function of the driving phase $\theta$. Unlike the numerical simulations, the analytical results in Eqs.~\eqref{eq:10} and \eqref{eq:11} describe the steady-state scattering for incident single plane-wave photons, where the transmission coefficient can approach $1$ in the ideal limit. For the case of $N=4n+3$ $(n\in\mathbb{N})$, the transmission coefficient $\tilde{T}_{b}$ exhibits the opposite phase dependence (by taking $\theta\rightarrow-\theta$) compared with the case of $N=4n+1$. This can be understood from the fact that the total phase difference between the two excitation paths, i.e. $\left|(k_{\mathrm{f}}N\ \mathrm{mod}\ 2\pi)-\theta\right|\ (k_{\mathrm{f}}N\in\left[-\pi,\pi\right)])$, remains unchanged for these two cases.

Note that the above analytical results are only valid when the width of the incident Gaussian wave packet is much larger than the coupling separation of the giant atom in the frequency dimension. For comparison, the dotted lines in Fig.~\ref{fig:4(a)} represent the situation where the width of the incident packet (i.e., $\sigma=20$) is smaller than the coupling separation (e.g., $N=33,35$). Numerical simulations reveal that in this case, the wave packet interacts with the giant atom through the two coupling points sequentially (rather than simultaneously), thus preventing the establishment of the self-interference effect and leading to significant discrepancies from the analytical predictions. Taking the coupling separation $N=4n+3$ as an example, Fig.~\ref{fig:4(b)} depicts the transmission coefficients as a function of the driving phase. Specifically, the incident plane wave is completely transmitted along $a$-FL when $\theta=\pi/2$. However, the incident plane wave is completely routed to $b$-FL when the driving phase is tuned to $\theta=-\pi/2$. This implies that the routing probability of our scheme can approach $1$ in the ideal limit. These results are also applicable for the incident wave coming from the right side of $a$-FL, where one should replace  $N \rightarrow -N$ and $\theta \rightarrow -\theta$.

\section{Discussion and Conclusion}

We would like to remark that here we use a superconducting quantum device as a three-level artificial atom, which allows for greater tunability of coupling strengths. For the adiabatic elimination process in Eq.~\eqref{eq:4}, it is necessary to satisfy the condition $g_{0}>g_{N}$ to ensure that the effective coupling strength $g_{0}^{\prime}=g_{N}$. Thanks to the high tunability of superconducting quantum circuits, achieving $g_{0}>g_{N}$ is within reach based on state-of-the-art experimental platforms~\cite{SQC.Murali,SQC.Wallraff}. Furthermore, we consider 400 frequency modes for each FL in our numerical simulations to avoid boundary effects. Experimentally, a proposed scheme involves a lithium niobate ring cavity can generate 900 uniformly distributed frequency modes in the infrared band through appropriate dynamic modulations, and a similar frequency spectrum can also be realized in the microwave regime~\cite{Broadband2019}. This demonstrates the feasibility of implementing our proposal under current experimental capabilities.

In conclusion, we have investigated the single-photon scattering problem of an effective giant atom coupled to a pair of synthetic lattices in the frequency dimension. We began with a model where a cyclic three-level artificial atom is coupled to a modulated multi-mode ring resonator. By coupling two transitions of the artificial atom to two frequency modes of the ring resonator, we have constructed an effective giant-atom model with two coupling points for each FL. Furthermore, the third transition is driven by an external field and thus introduces a tunable driving phase for the coupling points between the giant atom and the FLs. Compared with typical giant-atom experimental schemes in real space, where achieving a controllable coupling phase often requires the introduction of dynamically modulated magnetic flux~\cite{gong2024}, our model offers a simpler and more efficient way to control the coupling phase by tuning the driving phase. It not only allows for easier control of the coupling phase but also naturally ensures equal coupling strength between the giant atom and Lattices. Through both numerical and analytical solutions, we have demonstrated that the transmission direction of the incident photons in the frequency dimension can be controlled by tuning the driving phase. 

As a potential quantum communication device, our system offers additional intriguing phenomena, such as oscillating bound states, which enable interconversion between stationary and flying qubits. This interconversion, crucial for quantum information processing, can be readily achieved by tuning the driving phase in our scheme (See Appendix~\ref{sec:APPC} for more details). 
\begin{acknowledgments}
This work was supported by the National Natural Science Foundation of China (Grant No. 12274107 and No. 12174058), the Innovation Program for Quantum Science and Technology (Grant No. 2023ZD0300704), and the Research Funds of Hainan University [Grant No. KYQD(ZR)23010]. The authors thank L. Du and A. F. Kockum for valuable discussions. 
\end{acknowledgments}


\appendix
\begin{widetext}
\section{EFFECTIVE HAMILTONIAN AND EQUATIONS OF MOTION}
\label{sec:APPA}

The Hamiltonian of the system is given by 
\begin{eqnarray}
H & = & H_{0}+H_{\text{FL}}+H_{\text{Int}},\nonumber \\
H_{0} & = & \omega_{e}\left|e\right\rangle \left\langle e\right|+\omega_{f}\left|f\right\rangle \left\langle f\right|+\sum_{n}\omega_{n}\left(\hat{a}_{n}^{\dagger}\hat{a}_{n}+\hat{b}_{n}^{\dagger}\hat{b}_{n}\right),\nonumber \\
H_{\text{FL}} & = & -\sum_{m}J\left(\hat{a}_{m}^{\dagger}\hat{a}_{m+1}+\hat{b}_{m}^{\dagger}\hat{b}_{m+1}+\text{H.c.}\right),\nonumber \\
H_{\text{Int}} & = & g_{N}e^{i\Delta_{e}t}\left|g\right\rangle \left\langle e\right|\left(\hat{a}_{N}^{\dagger}+\hat{b}_{N}^{\dagger}\right)+g_{0}e^{i\Delta_{f}t}\left|g\right\rangle \left\langle f\right|\left(\hat{a}_{0}^{\dagger}+\hat{b}_{0}^{\dagger}\right)+\eta e^{i\theta}e^{i\Delta_{d}t}\left|e\right\rangle \left\langle f\right|+\text{H.c.}.\label{eq:A7}
\end{eqnarray}
Here $H_{0}$ represents the free Hamiltonian of the giant atom and frequency modes. $H_{\text{FL}}$ denotes the Hamiltonian of the 1D FLs. $H_{\text{Int}}$ is the interaction Hamiltonian describing the coupling between the giant atom, the driving field, and the 1D FLs. $\Delta_{e}=\omega_{e}-\omega_{N}\ (\Delta_{f}=\omega_{f}-\omega_{0})$ is the detuning between the modes $a_{0},b_{0}\ (a_{N},b_{N})$ and the transition $\left|g\right\rangle \leftrightarrow\left|e\right\rangle \ (\left|g\right\rangle \leftrightarrow\left|f\right\rangle )$. $\Delta_{d}=\omega_{e}-\omega_{f}-\omega_{d}$ is the detuning between the transition $\left|f\right\rangle \leftrightarrow\left|e\right\rangle$ and the external field. The state of the system in the single-excitation subspace is 
\begin{equation}
\left|\psi(t)\right\rangle =\sum_{m}\left[u_{m}(t)\hat{a}_{m}^{\dagger}+v_{m}(t)\hat{b}_{m}^{\dagger}\right]\left|0,g\right\rangle +\sum_{\beta=f,e}w_{\beta}(t)\left|0,\beta\right\rangle .
\end{equation}
In the rotating frame with respect to $U=e^{-i\left(\Delta_{f}\left|f\right\rangle \left\langle f\right|+\Delta_{e}\left|e\right\rangle \left\langle e\right|\right)t}$ and under the three-photon resonance condition $\omega_{d}=\omega_{N}-\omega_{0}$, which implies that $\Delta_{f}+\Delta_{d}-\Delta_{e}=0$, Eq.~\eqref{eq:A7} can be transformed into a time-independent form 
\begin{eqnarray}
H^{\prime} & = & H_{\text{FL}}-\Delta_{f}\left|f\right\rangle \left\langle f\right|-\Delta_{e}\left|e\right\rangle \left\langle e\right|+g_{N}a_{N}^{\dagger}\left|g\right\rangle \left\langle e\right|+g_{N}b_{N}^{\dagger}\left|g\right\rangle \left\langle e\right|\nonumber \\
 &  & +g_{0}b_{0}^{\dagger}\left|g\right\rangle \left\langle f\right|+g_{0}a_{0}^{\dagger}\left|g\right\rangle \left\langle f\right|+\eta e^{i\theta}\left|e\right\rangle \left\langle f\right|+\text{H.c.}.
\end{eqnarray}
By solving $i\frac{d}{dt}\left|\psi(t)\right\rangle =H^{\prime}\left|\psi(t)\right\rangle $ we obtain 
\begin{eqnarray}
i\dot{w}_{e} & = & \left(-i\gamma_{e}-\Delta_{e}\right)w_{e}+g_{N}(u_{m}+v_{m})\delta_{m,N}+\eta e^{i\theta}w_{f},\nonumber \\
i\dot{w}_{f} & = & \left(-i\gamma_{f}-\Delta_{f}\right)w_{f}+g_{0}(u_{m}+v_{m})\delta_{m,0}+\eta e^{-i\theta}w_{e},\nonumber \\
i\dot{u}_{m} & = & -J\left(u_{m+1}+u_{m-1}\right)+g_{N}w_{e}\delta_{m,N}+g_{0}w_{f}\delta_{m,0},\nonumber \\
i\dot{v}_{m} & = & -J\left(v_{m+1}+v_{m-1}\right)+g_{N}w_{e}\delta_{m,N}+g_{0}w_{f}\delta_{m,0},\label{eq:A:10}
\end{eqnarray}
where $u_{m}\left(v_{m}\right)$ is the probability amplitude of creating a photon in the $m$-th resonant mode across $a$-FL ($b$-FL), and $w_{e}\left(w_{f}\right)$ is the probability amplitude of the atom in the state $\left|e\right\rangle \left(\left|f\right\rangle \right)$. The terms $-i\gamma_{\beta}w_{\beta}$ ($\beta=e,f$) in Eq.~\eqref{eq:A:10} are introduced phenomenologically where $\gamma_{\beta}$ represent the intrinsic dissipation rate of state $\left|\beta\right\rangle $.

\section{SCATTERING STATES}

\label{APPB}

We calculate the scattering states for a giant atom coupled with two FLs. The single-excitation eigenstate is 
\begin{equation}
\left|\psi\right\rangle =\sum_{m}\left(u_{m}\hat{a}_{m}^{\dagger}+v_{m}\hat{b}_{m}^{\dagger}\right)\left|0,g\right\rangle +u_{e}\left|0,e\right\rangle .
\end{equation}
By solving the Schrödinger equation in frequency dimension $\hat{H}\left|\psi\right\rangle=E_{\mathrm{f}}\left|\psi\right\rangle$, we obtain 
\begin{eqnarray}
E_{\mathrm{f}}w_{e} & = & e^{i\theta}g\left(u_{0}+v_{0}\right)+g\left(u_{N}+v_{N}\right),\nonumber \\
E_{\mathrm{f}}u_{m} & = & -J\left(u_{m+1}+u_{m-1}\right)+e^{-i\theta}gw_{e}\delta_{m,0}+gw_{e}\delta_{m,N},\nonumber \\
E_{\mathrm{f}}v_{m} & = & -J\left(v_{m+1}+v_{m-1}\right)+e^{-i\theta}gw_{e}\delta_{m,0}+gw_{e}\delta_{m,N}.\label{eq:27}
\end{eqnarray}
The wave functions are given respectively by 
\begin{equation}
u_{m}=\begin{cases}
e^{ik_{\mathrm{fa}}m}+r_{a}e^{-ik_{\mathrm{fa}}m}, & m<0\\
l_{al}e^{-ik_{\mathrm{fa}}m}+l_{ar}e^{ik_{\mathrm{fa}}m}, & 0\leqslant m\leqslant N\\
t_{a}e^{ik_{\mathrm{fa}}m}, & m>N
\end{cases}\label{D3}
\end{equation}
and 
\begin{equation}
v_{m}=\begin{cases}
r_{b}e^{-ik_{\mathrm{fb}}m}, & m<0\\
l_{bl}e^{-ik_{\mathrm{fb}}m}+l_{br}e^{ik_{\mathrm{fb}}m}, & 0\leqslant m\leqslant N\\
t_{b}e^{ik_{\mathrm{fb}}m}. & m>N
\end{cases}\label{D4}
\end{equation}
Due to energy conservation in the process of photon absorption and re-emission in the frequency dimension, we assume that the wave propagating in both FLs has the same ``wave vector'', i.e., $k_{\mathrm{fa}}=k_{\mathrm{fb}}=k_{\mathrm{f}}$. By applying the continuity conditions at $m=0$ and $m=N$, the relationship of scattering amplitudes can be obtained: 
\begin{eqnarray}
1+r_{a} & = & l_{al}+l_{ar},\\
r_{b} & = & l_{bl}+l_{br},\\
t_{a}e^{ik_{\mathrm{f}}N} & = & l_{al}e^{-ik_{\mathrm{f}}N}+l_{ar}e^{ik_{\mathrm{f}}N},\\
t_{b}e^{ik_{\mathrm{f}}N} & = & l_{bl}e^{-ik_{\mathrm{f}}N}+l_{br}e^{ik_{\mathrm{f}}N}.
\end{eqnarray}
By substituting the Eqs. \eqref{D3} and \eqref{D4} into Eq.~\eqref{eq:27}
and using the continuity condition, we have 
\begin{eqnarray}
\left(4J^{2}\cos^{2}k_{\mathrm{f}}-g^{2}\right)\left(l_{al}+l_{ar}\right)-2J^{2}\cos k_{\mathrm{f}}\left(r_{a}e^{ik_{\mathrm{f}}}+l_{al}e^{-ik_{\mathrm{f}}}+l_{ar}e^{ik_{\mathrm{f}}}\right) & & \nonumber\\
-g^{2}r_{b}-g^{2}e^{-i\theta}t_{a}e^{ik_{\mathrm{f}}N}-g^{2}e^{-i\theta}t_{b}e^{ik_{\mathrm{f}}N}-2J^{2}\cos k_{\mathrm{f}}e^{-ik_{\mathrm{f}}} &= &0,\label{b9}
\end{eqnarray}
\begin{eqnarray}
\left(4J^{2}\cos^{2}k_{\mathrm{f}}-g^{2}\right)r_{b}-2J^{2}\cos k_{\mathrm{f}}\left(r_{b}e^{ik_{\mathrm{f}}}+l_{bl}e^{-ik_{\mathrm{f}}}+l_{br}e^{ik_{\mathrm{f}}}\right) & & \nonumber \\
-g^{2}\left(l_{al}+l_{ar}\right)-g^{2}e^{-i\theta}t_{a}e^{ik_{\mathrm{f}}N}-g^{2}e^{-i\theta}t_{b}e^{ik_{\mathrm{f}}N}&=&0,
\end{eqnarray}
\begin{eqnarray}
2J^{2}\cos k_{\mathrm{f}}\left[l_{al}e^{-ik_{\mathrm{f}}\left(N-1\right)}+l_{ar}e^{ik_{\mathrm{f}}\left(N-1\right)}+t_{a}e^{ik_{\mathrm{f}}\left(N+1\right)}\right] & & \nonumber \\
+e^{i\theta}g^{2}\left(l_{al}+l_{ar}\right)+e^{i\theta}g^{2}r_{b}+g^{2}t_{b}e^{ik_{\mathrm{f}}N}-\left(4J^{2}\cos^{2}k_{\mathrm{f}}-g^{2}\right)t_{a}e^{ik_{\mathrm{f}}N}&=&0,
\end{eqnarray}
\begin{eqnarray}
2J^{2}\cos k_{\mathrm{f}}\left[l_{bl}e^{-ik_{\mathrm{f}}\left(N-1\right)}+l_{br}e^{ik_{\mathrm{f}}\left(N-1\right)}+t_{b}e^{ik_{\mathrm{f}}\left(N+1\right)}\right] & & \nonumber \\
+e^{i\theta}g^{2}\left(l_{al}+l_{ar}\right)+e^{i\theta}g^{2}r_{b}+g^{2}t_{a}e^{ik_{\mathrm{f}}N}-\left(4J^{2}\cos^{2}k_{\mathrm{f}}-g^{2}\right)t_{b}e^{ik_{\mathrm{f}}N}&=&0.\label{b12}
\end{eqnarray}
By solving Eqs.~\eqref{b9}-\eqref{b12}, the scattering amplitudes
can be obtained

\begin{equation}
\left(\begin{array}{c}
r_{a}\\
l_{al}\\
l_{ar}\\
t_{a}\\
r_{b}\\
l_{bl}\\
l_{br}\\
t_{b}
\end{array}\right)=\left(\begin{array}{c}
-\frac{e^{ik_{\mathrm{f}}N}\left(\cos (k_{\mathrm{f}}N)+\cos\theta\right)}{2\left(\cos (k_{\mathrm{f}}N)\cos\theta+1\right)+i\xi}\\
-\frac{\frac{1}{2}e^{ik_{\mathrm{f}}N}\left(e^{i\theta}+e^{ik_{\mathrm{f}}N}\right)}{2\left(\cos (k_{\mathrm{f}}N)\cos\theta+1\right)+i\xi}\\
\frac{2\cos (k_{\mathrm{f}}N)\cos\theta+\frac{1}{2}\left[3-e^{i\left(k_{\mathrm{f}}N-\theta\right)}\right]+i\xi}{2\left(\cos (k_{\mathrm{f}}N)\cos\theta+1\right)+i\xi}\\
\frac{[1+\cos(k_{\mathrm{f}}N+\theta)]+i\xi}{2(\cos (k_{\mathrm{f}}N)\cos\theta+1)+i\xi}\\
-\frac{e^{ik_{\mathrm{f}}N}\left(\cos (k_{\mathrm{f}}N)+\cos\theta\right)}{2\left(\cos (k_{\mathrm{f}}N)\cos\theta+1\right)+i\xi}\\
-\frac{e^{ik_{\mathrm{f}}N}\cos\theta}{2\left(\cos (k_{\mathrm{f}}N)\cos\theta+1\right)+i\xi}\\
-\frac{\frac{1}{2}\left[1+e^{i(k_{\mathrm{f}}N-\theta)}\right]}{2\left(\cos (k_{\mathrm{f}}N)\cos\theta+1\right)+i\xi}\\
-\frac{\left[1+\cos(k_{\mathrm{f}}N-\theta)\right]}{2\left(\cos (k_{\mathrm{f}}N)\cos\theta+1\right)+i\xi}
\end{array}\right)
\end{equation}
with $\xi=2\sin (k_{\mathrm{f}}N)\cos\theta+J^{2}\sin(2k_{\mathrm{f}})/g^{2}$.

\section{OSCILLATING BOUND STATES FOR A NON-MARKOVIAN GIANT ATOM}

\label{sec:APPC}

One of the key objectives in implementing quantum communication is the ability to interconvert flying and stationary qubits~\cite{Quantum.Com}. Recent research has confirmed the oscillating bound states of bosonic field in systems where non-Markovian giant atoms are coupled with transmission lines. This findings demonstrate that non-Markovian giant atom can serve as effective tools for catching and releasing propagating bosonic field within the transmission lines, thereby highlighting their significant potential in quantum information applications~\cite{NM.Xu2024}.

In Sec.~\ref{subsec:3A}, the width of the initial wave packet is much larger than the coupling separation i.e., $\sigma\gg\left|N\right|$. Consequently, our results are confined to scenarios where the transmission time of the wave packet between two coupling points on the same frequency lattice (FL) is negligible. In this section, we examine the case where the coupling separation is larger than the excitation distribution width ($\sigma<\left|N\right|$).

\begin{figure}
\centering \subfigure{\includegraphics[width=0.4\linewidth]{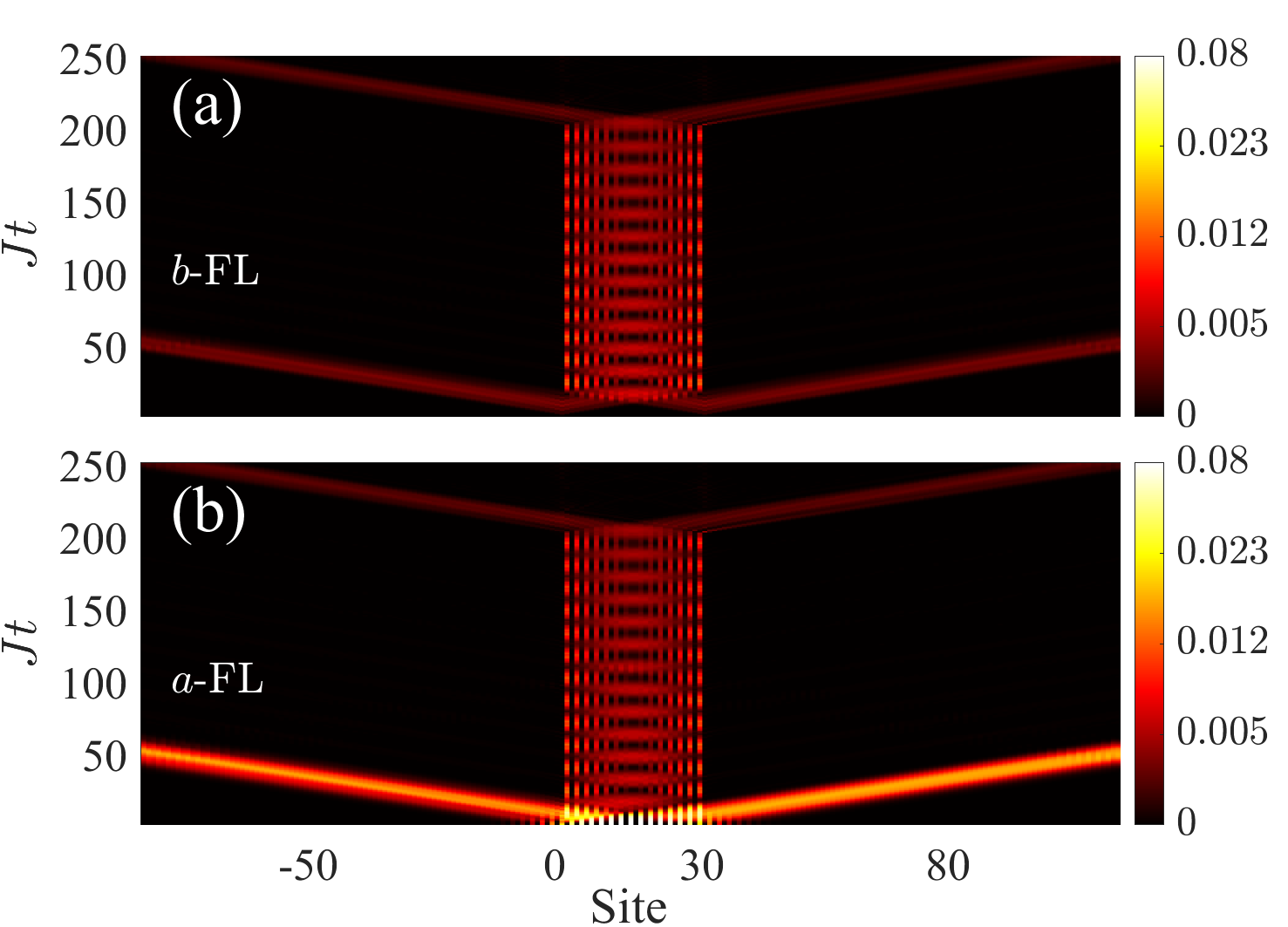}\label{fig:5(a)}}
\subfigure{\label{fig:5(b)}} \subfigure{\includegraphics[width=0.4\linewidth]{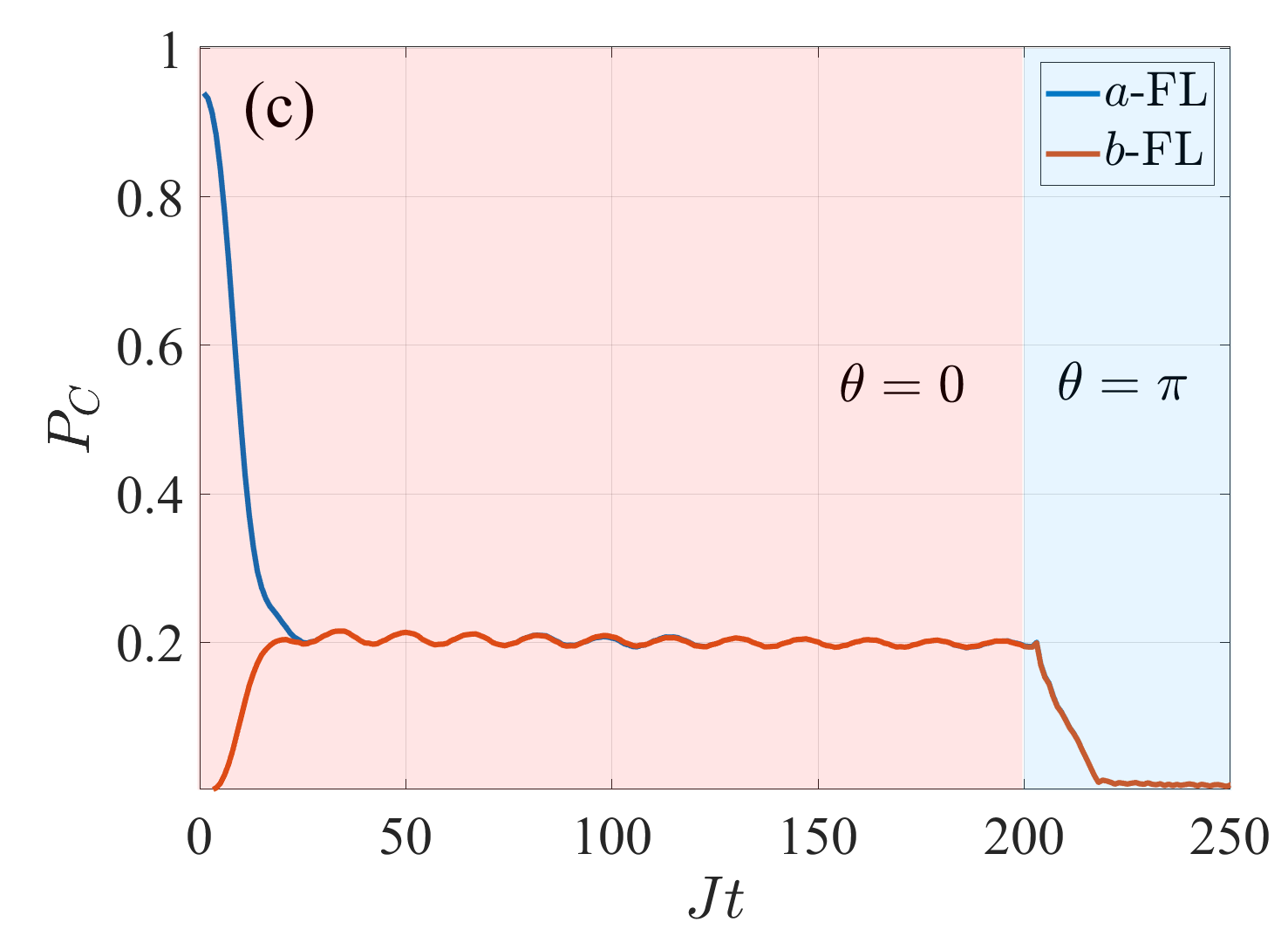}\label{fig:5(c)}}
\caption{(a) and (b) Field probability on the FL, with the initial states consisting of two Gaussian wave packets, centered at the coupling separation of the giant atom, propagating in opposite directions. (c) The excitation catch probabilities as a function of $Jt$. The parameters are $g_{0}=0.7J,\ g_{N}=4J,\ \eta=17.5J,\ \Delta_{f}=100J,\ N=30$ and $\sigma=16$. Release condition: $\theta=\pi$ when $Jt=200$.}
\end{figure}

Similar to the approach given in Ref.~\cite{NM.Xu2024}, we consider an initial state defined as $\left|\psi(0)\right\rangle =\left|\psi(0)\right\rangle _{L}+\left|\psi(0)\right\rangle _{R}$. This state is a superposition state of two counter-propagated Gaussian wave packets centered at $a_{N/2}$. We then numerically solve Eq.~\eqref{eq:5}, with 
\begin{eqnarray}
\left|\psi(0)\right\rangle _{L} & = & \frac{A}{\sqrt{2}}\sum_{m}\exp\left[-\frac{\left(m-\frac{N}{2}\right)^{2}}{2\sigma^{2}}+i\frac{\pi}{2}m\right]\hat{a}_{m}^{\dagger}\left|0,g\right\rangle ,\\
\left|\psi(0)\right\rangle _{R} & = & \frac{A}{\sqrt{2}}\sum_{m}\exp\left[-\frac{\left(m-\frac{N}{2}\right)^{2}}{2\sigma^{2}}-i\frac{\pi}{2}m\right]\hat{a}_{m}^{\dagger}\left|0,g\right\rangle .
\end{eqnarray}
Here, $\left|\psi(0)\right\rangle _{L}$ ($\left|\psi(0)\right\rangle _{R}$) represents Gaussian wave packet propagating to the left (right) in the $a$-FL. In Figs.~\ref{fig:5(a)}~and~\ref{fig:5(b)}, we plot the time evolution of excitation probability. Initially, the Gaussian wave packets propagate from their respective starting position and interact with the giant atom via the two coupling sites. When $Jt>25$, the incident Gaussian wave packet is routed to the $b$-FL, with a few excitations scattered outside the coupling region between the coupling sites $a_{0}$ and $a_{N}$ at the initial time. This scattering occurs because the giant atom requires a finite time to establish interactions among different coupling sites. In the time separation $25<Jt<200$, photons form an oscillating bound state between two coupling sites $a_{0}\ (b_{0})$ and $a_{N}\ (b_{N})$. After $Jt=200$, we initiate the photon release process by tuning the driving phase $\theta=\pi$. We define the excitation catch probabilities between the two coupling sites as $P_{C}$. As depicted in Fig.~\ref{fig:5(c)}, the excitation capture probabilities between the two coupling sites on each FL are approximately $P_{C}=20\%$. Subsequently, the excitation is released from the coupling region and routed to different sides of the FL when $\theta=\pi$. These results, including the formation of oscillating bound states and the controlled release of photons, validate the ability of this non-Markovian giant atom to interconvert stationary and flying qubits.
\end{widetext}

\bibliographystyle{apsrev4-2}
\bibliography{ref}

\begin{thebibliography}{74}%
\makeatletter
\providecommand \@ifxundefined [1]{%
 \@ifx{#1\undefined}
}%
\providecommand \@ifnum [1]{%
 \ifnum #1\expandafter \@firstoftwo
 \else \expandafter \@secondoftwo
 \fi
}%
\providecommand \@ifx [1]{%
 \ifx #1\expandafter \@firstoftwo
 \else \expandafter \@secondoftwo
 \fi
}%
\providecommand \natexlab [1]{#1}%
\providecommand \enquote  [1]{``#1''}%
\providecommand \bibnamefont  [1]{#1}%
\providecommand \bibfnamefont [1]{#1}%
\providecommand \citenamefont [1]{#1}%
\providecommand \href@noop [0]{\@secondoftwo}%
\providecommand \href [0]{\begingroup \@sanitize@url \@href}%
\providecommand \@href[1]{\@@startlink{#1}\@@href}%
\providecommand \@@href[1]{\endgroup#1\@@endlink}%
\providecommand \@sanitize@url [0]{\catcode `\\12\catcode `\$12\catcode `\&12\catcode `\#12\catcode `\^12\catcode `\_12\catcode `\%12\relax}%
\providecommand \@@startlink[1]{}%
\providecommand \@@endlink[0]{}%
\providecommand \url  [0]{\begingroup\@sanitize@url \@url }%
\providecommand \@url [1]{\endgroup\@href {#1}{\urlprefix }}%
\providecommand \urlprefix  [0]{URL }%
\providecommand \Eprint [0]{\href }%
\providecommand \doibase [0]{https://doi.org/}%
\providecommand \selectlanguage [0]{\@gobble}%
\providecommand \bibinfo  [0]{\@secondoftwo}%
\providecommand \bibfield  [0]{\@secondoftwo}%
\providecommand \translation [1]{[#1]}%
\providecommand \BibitemOpen [0]{}%
\providecommand \bibitemStop [0]{}%
\providecommand \bibitemNoStop [0]{.\EOS\space}%
\providecommand \EOS [0]{\spacefactor3000\relax}%
\providecommand \BibitemShut  [1]{\csname bibitem#1\endcsname}%
\let\auto@bib@innerbib\@empty
\bibitem [{\citenamefont {DiVincenzo}(2000)}]{Quantum.Com}%
  \BibitemOpen
  \bibfield  {author} {\bibinfo {author} {\bibfnamefont {D.~P.}\ \bibnamefont {DiVincenzo}},\ }\href {https://doi.org/https://doi.org/10.1002/1521-3978(200009)48:9/11<771::AID-PROP771>3.0.CO;2-E} {\bibfield  {journal} {\bibinfo  {journal} {Fortschritte der Physik}\ }\textbf {\bibinfo {volume} {48}},\ \bibinfo {pages} {771} (\bibinfo {year} {2000})}\BibitemShut {NoStop}%
\bibitem [{\citenamefont {Cirac}\ \emph {et~al.}(1997)\citenamefont {Cirac}, \citenamefont {Zoller}, \citenamefont {Kimble},\ and\ \citenamefont {Mabuchi}}]{Cirac.1997}%
  \BibitemOpen
  \bibfield  {author} {\bibinfo {author} {\bibfnamefont {J.~I.}\ \bibnamefont {Cirac}}, \bibinfo {author} {\bibfnamefont {P.}~\bibnamefont {Zoller}}, \bibinfo {author} {\bibfnamefont {H.~J.}\ \bibnamefont {Kimble}},\ and\ \bibinfo {author} {\bibfnamefont {H.}~\bibnamefont {Mabuchi}},\ }\href {https://doi.org/10.1103/PhysRevLett.78.3221} {\bibfield  {journal} {\bibinfo  {journal} {Phys. Rev. Lett.}\ }\textbf {\bibinfo {volume} {78}},\ \bibinfo {pages} {3221} (\bibinfo {year} {1997})}\BibitemShut {NoStop}%
\bibitem [{\citenamefont {Kimble}(2008)}]{kimble2008}%
  \BibitemOpen
  \bibfield  {author} {\bibinfo {author} {\bibfnamefont {H.~J.}\ \bibnamefont {Kimble}},\ }\href {https://doi.org/10.1038/nature07127} {\bibfield  {journal} {\bibinfo  {journal} {Nature (London)}\ }\textbf {\bibinfo {volume} {453}},\ \bibinfo {pages} {1023} (\bibinfo {year} {2008})}\BibitemShut {NoStop}%
\bibitem [{\citenamefont {Zhou}\ \emph {et~al.}(2013)\citenamefont {Zhou}, \citenamefont {Yang}, \citenamefont {Li},\ and\ \citenamefont {Sun}}]{zhou2013}%
  \BibitemOpen
  \bibfield  {author} {\bibinfo {author} {\bibfnamefont {L.}~\bibnamefont {Zhou}}, \bibinfo {author} {\bibfnamefont {L.-P.}\ \bibnamefont {Yang}}, \bibinfo {author} {\bibfnamefont {Y.}~\bibnamefont {Li}},\ and\ \bibinfo {author} {\bibfnamefont {C.~P.}\ \bibnamefont {Sun}},\ }\href {https://doi.org/10.1103/PhysRevLett.111.103604} {\bibfield  {journal} {\bibinfo  {journal} {Phys. Rev. Lett.}\ }\textbf {\bibinfo {volume} {111}},\ \bibinfo {pages} {103604} (\bibinfo {year} {2013})}\BibitemShut {NoStop}%
\bibitem [{\citenamefont {Lee}\ \emph {et~al.}(2022)\citenamefont {Lee}, \citenamefont {Bersin}, \citenamefont {Dahlberg}, \citenamefont {Wehner},\ and\ \citenamefont {Englund}}]{lee2022}%
  \BibitemOpen
  \bibfield  {author} {\bibinfo {author} {\bibfnamefont {Y.}~\bibnamefont {Lee}}, \bibinfo {author} {\bibfnamefont {E.}~\bibnamefont {Bersin}}, \bibinfo {author} {\bibfnamefont {A.}~\bibnamefont {Dahlberg}}, \bibinfo {author} {\bibfnamefont {S.}~\bibnamefont {Wehner}},\ and\ \bibinfo {author} {\bibfnamefont {D.}~\bibnamefont {Englund}},\ }\href {https://doi.org/10.1038/s41534-022-00582-8} {\bibfield  {journal} {\bibinfo  {journal} {npj Quantum Inf.}\ }\textbf {\bibinfo {volume} {8}},\ \bibinfo {pages} {1} (\bibinfo {year} {2022})}\BibitemShut {NoStop}%
\bibitem [{\citenamefont {Zhang}\ \emph {et~al.}(2022)\citenamefont {Zhang}, \citenamefont {Zhu}, \citenamefont {Chen}, \citenamefont {Peng}, \citenamefont {Yin}, \citenamefont {Yang}, \citenamefont {Zhao}, \citenamefont {Lu}, \citenamefont {Chai}, \citenamefont {Xiong},\ and\ \citenamefont {Tan}}]{zhang2022}%
  \BibitemOpen
  \bibfield  {author} {\bibinfo {author} {\bibfnamefont {Y.}~\bibnamefont {Zhang}}, \bibinfo {author} {\bibfnamefont {Z.}~\bibnamefont {Zhu}}, \bibinfo {author} {\bibfnamefont {K.}~\bibnamefont {Chen}}, \bibinfo {author} {\bibfnamefont {Z.}~\bibnamefont {Peng}}, \bibinfo {author} {\bibfnamefont {W.}~\bibnamefont {Yin}}, \bibinfo {author} {\bibfnamefont {Y.}~\bibnamefont {Yang}}, \bibinfo {author} {\bibfnamefont {Y.}~\bibnamefont {Zhao}}, \bibinfo {author} {\bibfnamefont {Z.}~\bibnamefont {Lu}}, \bibinfo {author} {\bibfnamefont {Y.}~\bibnamefont {Chai}}, \bibinfo {author} {\bibfnamefont {Z.}~\bibnamefont {Xiong}},\ and\ \bibinfo {author} {\bibfnamefont {L.}~\bibnamefont {Tan}},\ }\href {https://www.frontiersin.org/journals/physics/articles/10.3389/fphy.2022.1054299/full} {\bibfield  {journal} {\bibinfo  {journal} {Frontiers in Physics}\ }\textbf {\bibinfo {volume} {10}},\ \bibinfo {pages} {1054299} (\bibinfo {year} {2022})}\BibitemShut {NoStop}%
\bibitem [{\citenamefont {Wang}\ \emph {et~al.}(2021{\natexlab{a}})\citenamefont {Wang}, \citenamefont {Ma},\ and\ \citenamefont {Cheng}}]{Cheng2021}%
  \BibitemOpen
  \bibfield  {author} {\bibinfo {author} {\bibfnamefont {C.}~\bibnamefont {Wang}}, \bibinfo {author} {\bibfnamefont {X.-S.}\ \bibnamefont {Ma}},\ and\ \bibinfo {author} {\bibfnamefont {M.-T.}\ \bibnamefont {Cheng}},\ }\href {https://doi.org/10.1364/OE.444096} {\bibfield  {journal} {\bibinfo  {journal} {Opt. Express}\ }\textbf {\bibinfo {volume} {29}},\ \bibinfo {pages} {40116} (\bibinfo {year} {2021}{\natexlab{a}})}\BibitemShut {NoStop}%
\bibitem [{\citenamefont {Li}\ \emph {et~al.}(2022)\citenamefont {Li}, \citenamefont {Dong}, \citenamefont {Zhang},\ and\ \citenamefont {Wu}}]{flying2022}%
  \BibitemOpen
  \bibfield  {author} {\bibinfo {author} {\bibfnamefont {W.}~\bibnamefont {Li}}, \bibinfo {author} {\bibfnamefont {X.}~\bibnamefont {Dong}}, \bibinfo {author} {\bibfnamefont {G.}~\bibnamefont {Zhang}},\ and\ \bibinfo {author} {\bibfnamefont {R.-B.}\ \bibnamefont {Wu}},\ }\href {https://doi.org/10.1103/PhysRevB.106.134305} {\bibfield  {journal} {\bibinfo  {journal} {Phys. Rev. B}\ }\textbf {\bibinfo {volume} {106}},\ \bibinfo {pages} {134305} (\bibinfo {year} {2022})}\BibitemShut {NoStop}%
\bibitem [{\citenamefont {Li}\ \emph {et~al.}(2024{\natexlab{a}})\citenamefont {Li}, \citenamefont {Sun}, \citenamefont {Liu}, \citenamefont {Li},\ and\ \citenamefont {Wu}}]{flying2024}%
  \BibitemOpen
  \bibfield  {author} {\bibinfo {author} {\bibfnamefont {W.}~\bibnamefont {Li}}, \bibinfo {author} {\bibfnamefont {H.}~\bibnamefont {Sun}}, \bibinfo {author} {\bibfnamefont {Y.}~\bibnamefont {Liu}}, \bibinfo {author} {\bibfnamefont {T.}~\bibnamefont {Li}},\ and\ \bibinfo {author} {\bibfnamefont {R.-B.}\ \bibnamefont {Wu}},\ }\href {https://doi.org/10.34133/adi.0059} {\bibfield  {journal} {\bibinfo  {journal} {Adv. Devices Instrum.}\ }\textbf {\bibinfo {volume} {5}},\ \bibinfo {pages} {0059} (\bibinfo {year} {2024}{\natexlab{a}})}\BibitemShut {NoStop}%
\bibitem [{\citenamefont {Kockum}(2021)}]{Giant.atom.summarize}%
  \BibitemOpen
  \bibfield  {author} {\bibinfo {author} {\bibfnamefont {A.~F.}\ \bibnamefont {Kockum}},\ }in\ \href {https://link.springer.com/chapter/10.1007/978-981-15-5191-8_12} {\emph {\bibinfo {booktitle} {Mathematics for Industry}}}\ (\bibinfo {organization} {Springer Singapore},\ \bibinfo {year} {2021})\ pp.\ \bibinfo {pages} {125--146}\BibitemShut {NoStop}%
\bibitem [{\citenamefont {Gustafsson}\ \emph {et~al.}(2014)\citenamefont {Gustafsson}, \citenamefont {Aref}, \citenamefont {Kockum}, \citenamefont {Ekström}, \citenamefont {Johansson},\ and\ \citenamefont {Delsing}}]{SAW.Gustafsson}%
  \BibitemOpen
  \bibfield  {author} {\bibinfo {author} {\bibfnamefont {M.~V.}\ \bibnamefont {Gustafsson}}, \bibinfo {author} {\bibfnamefont {T.}~\bibnamefont {Aref}}, \bibinfo {author} {\bibfnamefont {A.~F.}\ \bibnamefont {Kockum}}, \bibinfo {author} {\bibfnamefont {M.~K.}\ \bibnamefont {Ekström}}, \bibinfo {author} {\bibfnamefont {G.}~\bibnamefont {Johansson}},\ and\ \bibinfo {author} {\bibfnamefont {P.}~\bibnamefont {Delsing}},\ }\href {https://doi.org/10.1126/science.1257219} {\bibfield  {journal} {\bibinfo  {journal} {Science}\ }\textbf {\bibinfo {volume} {346}},\ \bibinfo {pages} {207} (\bibinfo {year} {2014})}\BibitemShut {NoStop}%
\bibitem [{\citenamefont {Frisk~Kockum}\ \emph {et~al.}(2014)\citenamefont {Frisk~Kockum}, \citenamefont {Delsing},\ and\ \citenamefont {Johansson}}]{meandering.Kockum}%
  \BibitemOpen
  \bibfield  {author} {\bibinfo {author} {\bibfnamefont {A.}~\bibnamefont {Frisk~Kockum}}, \bibinfo {author} {\bibfnamefont {P.}~\bibnamefont {Delsing}},\ and\ \bibinfo {author} {\bibfnamefont {G.}~\bibnamefont {Johansson}},\ }\href {https://doi.org/10.1103/PhysRevA.90.013837} {\bibfield  {journal} {\bibinfo  {journal} {Phys. Rev. A}\ }\textbf {\bibinfo {volume} {90}},\ \bibinfo {pages} {013837} (\bibinfo {year} {2014})}\BibitemShut {NoStop}%
\bibitem [{\citenamefont {Chen}\ and\ \citenamefont {Kockum}(2024)}]{chenGZ2024}%
  \BibitemOpen
  \bibfield  {author} {\bibinfo {author} {\bibfnamefont {G.}~\bibnamefont {Chen}}\ and\ \bibinfo {author} {\bibfnamefont {A.~F.}\ \bibnamefont {Kockum}},\ }\href {https://arxiv.org/abs/2406.13678} {\bibfield  {journal} {\bibinfo  {journal} {arXiv:2406.13678}\ } (\bibinfo {year} {2024})}\BibitemShut {NoStop}%
\bibitem [{\citenamefont {Xu}\ and\ \citenamefont {Guo}(2024)}]{NM.Xu2024}%
  \BibitemOpen
  \bibfield  {author} {\bibinfo {author} {\bibfnamefont {L.}~\bibnamefont {Xu}}\ and\ \bibinfo {author} {\bibfnamefont {L.}~\bibnamefont {Guo}},\ }\href {https://doi.org/10.1088/1367-2630/ad18ed} {\bibfield  {journal} {\bibinfo  {journal} {New J. Phys.}\ }\textbf {\bibinfo {volume} {26}},\ \bibinfo {pages} {013025} (\bibinfo {year} {2024})}\BibitemShut {NoStop}%
\bibitem [{\citenamefont {Andersson}\ \emph {et~al.}(2019{\natexlab{a}})\citenamefont {Andersson}, \citenamefont {Suri}, \citenamefont {Guo}, \citenamefont {Aref},\ and\ \citenamefont {Delsing}}]{SAW.Andersson2019}%
  \BibitemOpen
  \bibfield  {author} {\bibinfo {author} {\bibfnamefont {G.}~\bibnamefont {Andersson}}, \bibinfo {author} {\bibfnamefont {B.}~\bibnamefont {Suri}}, \bibinfo {author} {\bibfnamefont {L.}~\bibnamefont {Guo}}, \bibinfo {author} {\bibfnamefont {T.}~\bibnamefont {Aref}},\ and\ \bibinfo {author} {\bibfnamefont {P.}~\bibnamefont {Delsing}},\ }\href {https://doi.org/10.1038/s41567-019-0605-6} {\bibfield  {journal} {\bibinfo  {journal} {Nat. Phys.}\ }\textbf {\bibinfo {volume} {15}},\ \bibinfo {pages} {1123} (\bibinfo {year} {2019}{\natexlab{a}})}\BibitemShut {NoStop}%
\bibitem [{\citenamefont {Andersson}\ \emph {et~al.}(2020)\citenamefont {Andersson}, \citenamefont {Ekstr\"om},\ and\ \citenamefont {Delsing}}]{SAW.Andersson2020}%
  \BibitemOpen
  \bibfield  {author} {\bibinfo {author} {\bibfnamefont {G.}~\bibnamefont {Andersson}}, \bibinfo {author} {\bibfnamefont {M.~K.}\ \bibnamefont {Ekstr\"om}},\ and\ \bibinfo {author} {\bibfnamefont {P.}~\bibnamefont {Delsing}},\ }\href {https://doi.org/10.1103/PhysRevLett.124.240402} {\bibfield  {journal} {\bibinfo  {journal} {Phys. Rev. Lett.}\ }\textbf {\bibinfo {volume} {124}},\ \bibinfo {pages} {240402} (\bibinfo {year} {2020})}\BibitemShut {NoStop}%
\bibitem [{\citenamefont {Guo}\ \emph {et~al.}(2017)\citenamefont {Guo}, \citenamefont {Grimsmo}, \citenamefont {Kockum}, \citenamefont {Pletyukhov},\ and\ \citenamefont {Johansson}}]{SAW.Guo}%
  \BibitemOpen
  \bibfield  {author} {\bibinfo {author} {\bibfnamefont {L.}~\bibnamefont {Guo}}, \bibinfo {author} {\bibfnamefont {A.}~\bibnamefont {Grimsmo}}, \bibinfo {author} {\bibfnamefont {A.~F.}\ \bibnamefont {Kockum}}, \bibinfo {author} {\bibfnamefont {M.}~\bibnamefont {Pletyukhov}},\ and\ \bibinfo {author} {\bibfnamefont {G.}~\bibnamefont {Johansson}},\ }\href {https://doi.org/10.1103/PhysRevA.95.053821} {\bibfield  {journal} {\bibinfo  {journal} {Phys. Rev. A}\ }\textbf {\bibinfo {volume} {95}},\ \bibinfo {pages} {053821} (\bibinfo {year} {2017})}\BibitemShut {NoStop}%
\bibitem [{\citenamefont {Manenti}\ \emph {et~al.}(2017)\citenamefont {Manenti}, \citenamefont {Kockum}, \citenamefont {Patterson}, \citenamefont {Behrle}, \citenamefont {Rahamim}, \citenamefont {Tancredi}, \citenamefont {Nori},\ and\ \citenamefont {Leek}}]{SAW.Manenti}%
  \BibitemOpen
  \bibfield  {author} {\bibinfo {author} {\bibfnamefont {R.}~\bibnamefont {Manenti}}, \bibinfo {author} {\bibfnamefont {A.~F.}\ \bibnamefont {Kockum}}, \bibinfo {author} {\bibfnamefont {A.}~\bibnamefont {Patterson}}, \bibinfo {author} {\bibfnamefont {T.}~\bibnamefont {Behrle}}, \bibinfo {author} {\bibfnamefont {J.}~\bibnamefont {Rahamim}}, \bibinfo {author} {\bibfnamefont {G.}~\bibnamefont {Tancredi}}, \bibinfo {author} {\bibfnamefont {F.}~\bibnamefont {Nori}},\ and\ \bibinfo {author} {\bibfnamefont {P.~J.}\ \bibnamefont {Leek}},\ }\href {https://doi.org/10.1038/s41467-017-01063-9} {\bibfield  {journal} {\bibinfo  {journal} {Nat. Commun.}\ }\textbf {\bibinfo {volume} {8}},\ \bibinfo {pages} {975} (\bibinfo {year} {2017})}\BibitemShut {NoStop}%
\bibitem [{\citenamefont {Noguchi}\ \emph {et~al.}(2017)\citenamefont {Noguchi}, \citenamefont {Yamazaki}, \citenamefont {Tabuchi},\ and\ \citenamefont {Nakamura}}]{SAW.Noguchi}%
  \BibitemOpen
  \bibfield  {author} {\bibinfo {author} {\bibfnamefont {A.}~\bibnamefont {Noguchi}}, \bibinfo {author} {\bibfnamefont {R.}~\bibnamefont {Yamazaki}}, \bibinfo {author} {\bibfnamefont {Y.}~\bibnamefont {Tabuchi}},\ and\ \bibinfo {author} {\bibfnamefont {Y.}~\bibnamefont {Nakamura}},\ }\href {https://doi.org/10.1103/PhysRevLett.119.180505} {\bibfield  {journal} {\bibinfo  {journal} {Phys. Rev. Lett.}\ }\textbf {\bibinfo {volume} {119}},\ \bibinfo {pages} {180505} (\bibinfo {year} {2017})}\BibitemShut {NoStop}%
\bibitem [{\citenamefont {Qiu}\ \emph {et~al.}(2023)\citenamefont {Qiu}, \citenamefont {Wu},\ and\ \citenamefont {Lü}}]{SAW.Qiu}%
  \BibitemOpen
  \bibfield  {author} {\bibinfo {author} {\bibfnamefont {Q.-Y.}\ \bibnamefont {Qiu}}, \bibinfo {author} {\bibfnamefont {Y.}~\bibnamefont {Wu}},\ and\ \bibinfo {author} {\bibfnamefont {X.-Y.}\ \bibnamefont {Lü}},\ }\href {https://doi.org/10.1007/s11433-022-1990-x} {\bibfield  {journal} {\bibinfo  {journal} {Sci. China Phys. Mech. Astron.}\ }\textbf {\bibinfo {volume} {66}},\ \bibinfo {pages} {224212} (\bibinfo {year} {2023})}\BibitemShut {NoStop}%
\bibitem [{\citenamefont {Kannan}\ \emph {et~al.}(2020)\citenamefont {Kannan}, \citenamefont {Ruckriegel}, \citenamefont {Campbell}, \citenamefont {Frisk~Kockum}, \citenamefont {Braumüller}, \citenamefont {Kim}, \citenamefont {Kjaergaard}, \citenamefont {Krantz}, \citenamefont {Melville}, \citenamefont {Niedzielski}, \citenamefont {Vepsäläinen}, \citenamefont {Winik}, \citenamefont {Yoder}, \citenamefont {Nori}, \citenamefont {Orlando}, \citenamefont {Gustavsson},\ and\ \citenamefont {Oliver}}]{meandering.Kannan}%
  \BibitemOpen
  \bibfield  {author} {\bibinfo {author} {\bibfnamefont {B.}~\bibnamefont {Kannan}}, \bibinfo {author} {\bibfnamefont {M.~J.}\ \bibnamefont {Ruckriegel}}, \bibinfo {author} {\bibfnamefont {D.~L.}\ \bibnamefont {Campbell}}, \bibinfo {author} {\bibfnamefont {A.}~\bibnamefont {Frisk~Kockum}}, \bibinfo {author} {\bibfnamefont {J.}~\bibnamefont {Braumüller}}, \bibinfo {author} {\bibfnamefont {D.~K.}\ \bibnamefont {Kim}}, \bibinfo {author} {\bibfnamefont {M.}~\bibnamefont {Kjaergaard}}, \bibinfo {author} {\bibfnamefont {P.}~\bibnamefont {Krantz}}, \bibinfo {author} {\bibfnamefont {A.}~\bibnamefont {Melville}}, \bibinfo {author} {\bibfnamefont {B.~M.}\ \bibnamefont {Niedzielski}}, \bibinfo {author} {\bibfnamefont {A.}~\bibnamefont {Vepsäläinen}}, \bibinfo {author} {\bibfnamefont {R.}~\bibnamefont {Winik}}, \bibinfo {author} {\bibfnamefont {J.~L.}\ \bibnamefont {Yoder}}, \bibinfo {author} {\bibfnamefont {F.}~\bibnamefont {Nori}}, \bibinfo {author} {\bibfnamefont {T.~P.}\ \bibnamefont {Orlando}}, \bibinfo {author}
  {\bibfnamefont {S.}~\bibnamefont {Gustavsson}},\ and\ \bibinfo {author} {\bibfnamefont {W.~D.}\ \bibnamefont {Oliver}},\ }\href {https://doi.org/10.1038/s41586-020-2529-9} {\bibfield  {journal} {\bibinfo  {journal} {Nature (London)}\ }\textbf {\bibinfo {volume} {583}},\ \bibinfo {pages} {775} (\bibinfo {year} {2020})}\BibitemShut {NoStop}%
\bibitem [{\citenamefont {Vadiraj}\ \emph {et~al.}(2021)\citenamefont {Vadiraj}, \citenamefont {Ask}, \citenamefont {McConkey}, \citenamefont {Nsanzineza}, \citenamefont {Chang}, \citenamefont {Kockum},\ and\ \citenamefont {Wilson}}]{meandering.Vadiraj}%
  \BibitemOpen
  \bibfield  {author} {\bibinfo {author} {\bibfnamefont {A.~M.}\ \bibnamefont {Vadiraj}}, \bibinfo {author} {\bibfnamefont {A.}~\bibnamefont {Ask}}, \bibinfo {author} {\bibfnamefont {T.~G.}\ \bibnamefont {McConkey}}, \bibinfo {author} {\bibfnamefont {I.}~\bibnamefont {Nsanzineza}}, \bibinfo {author} {\bibfnamefont {C.~W.~S.}\ \bibnamefont {Chang}}, \bibinfo {author} {\bibfnamefont {A.~F.}\ \bibnamefont {Kockum}},\ and\ \bibinfo {author} {\bibfnamefont {C.~M.}\ \bibnamefont {Wilson}},\ }\href {https://doi.org/10.1103/PhysRevA.103.023710} {\bibfield  {journal} {\bibinfo  {journal} {Phys. Rev. A}\ }\textbf {\bibinfo {volume} {103}},\ \bibinfo {pages} {023710} (\bibinfo {year} {2021})}\BibitemShut {NoStop}%
\bibitem [{\citenamefont {Du}\ \emph {et~al.}(2023{\natexlab{a}})\citenamefont {Du}, \citenamefont {Chen}, \citenamefont {Zhang}, \citenamefont {Li},\ and\ \citenamefont {Wu}}]{Du_2023}%
  \BibitemOpen
  \bibfield  {author} {\bibinfo {author} {\bibfnamefont {L.}~\bibnamefont {Du}}, \bibinfo {author} {\bibfnamefont {Y.-T.}\ \bibnamefont {Chen}}, \bibinfo {author} {\bibfnamefont {Y.}~\bibnamefont {Zhang}}, \bibinfo {author} {\bibfnamefont {Y.}~\bibnamefont {Li}},\ and\ \bibinfo {author} {\bibfnamefont {J.-H.}\ \bibnamefont {Wu}},\ }\href {https://doi.org/10.1088/2058-9565/ace54c} {\bibfield  {journal} {\bibinfo  {journal} {Quantum Sci. Technol.}\ }\textbf {\bibinfo {volume} {8}},\ \bibinfo {pages} {045010} (\bibinfo {year} {2023}{\natexlab{a}})}\BibitemShut {NoStop}%
\bibitem [{\citenamefont {Wang}\ \emph {et~al.}(2021{\natexlab{b}})\citenamefont {Wang}, \citenamefont {Liu}, \citenamefont {Kockum}, \citenamefont {Li},\ and\ \citenamefont {Nori}}]{Wang2021}%
  \BibitemOpen
  \bibfield  {author} {\bibinfo {author} {\bibfnamefont {X.}~\bibnamefont {Wang}}, \bibinfo {author} {\bibfnamefont {T.}~\bibnamefont {Liu}}, \bibinfo {author} {\bibfnamefont {A.~F.}\ \bibnamefont {Kockum}}, \bibinfo {author} {\bibfnamefont {H.-R.}\ \bibnamefont {Li}},\ and\ \bibinfo {author} {\bibfnamefont {F.}~\bibnamefont {Nori}},\ }\href {https://doi.org/10.1103/PhysRevLett.126.043602} {\bibfield  {journal} {\bibinfo  {journal} {Phys. Rev. Lett.}\ }\textbf {\bibinfo {volume} {126}},\ \bibinfo {pages} {043602} (\bibinfo {year} {2021}{\natexlab{b}})}\BibitemShut {NoStop}%
\bibitem [{\citenamefont {Carollo}\ \emph {et~al.}(2020)\citenamefont {Carollo}, \citenamefont {Cilluffo},\ and\ \citenamefont {Ciccarello}}]{DFI.Carollo}%
  \BibitemOpen
  \bibfield  {author} {\bibinfo {author} {\bibfnamefont {A.}~\bibnamefont {Carollo}}, \bibinfo {author} {\bibfnamefont {D.}~\bibnamefont {Cilluffo}},\ and\ \bibinfo {author} {\bibfnamefont {F.}~\bibnamefont {Ciccarello}},\ }\href {https://doi.org/10.1103/PhysRevResearch.2.043184} {\bibfield  {journal} {\bibinfo  {journal} {Phys. Rev. Res.}\ }\textbf {\bibinfo {volume} {2}},\ \bibinfo {pages} {043184} (\bibinfo {year} {2020})}\BibitemShut {NoStop}%
\bibitem [{\citenamefont {Kockum}\ \emph {et~al.}(2018)\citenamefont {Kockum}, \citenamefont {Johansson},\ and\ \citenamefont {Nori}}]{DFI.Kockum}%
  \BibitemOpen
  \bibfield  {author} {\bibinfo {author} {\bibfnamefont {A.~F.}\ \bibnamefont {Kockum}}, \bibinfo {author} {\bibfnamefont {G.}~\bibnamefont {Johansson}},\ and\ \bibinfo {author} {\bibfnamefont {F.}~\bibnamefont {Nori}},\ }\href {https://doi.org/10.1103/PhysRevLett.120.140404} {\bibfield  {journal} {\bibinfo  {journal} {Phys. Rev. Lett.}\ }\textbf {\bibinfo {volume} {120}},\ \bibinfo {pages} {140404} (\bibinfo {year} {2018})}\BibitemShut {NoStop}%
\bibitem [{\citenamefont {Du}\ \emph {et~al.}(2023{\natexlab{b}})\citenamefont {Du}, \citenamefont {Guo},\ and\ \citenamefont {Li}}]{DFI.Lei}%
  \BibitemOpen
  \bibfield  {author} {\bibinfo {author} {\bibfnamefont {L.}~\bibnamefont {Du}}, \bibinfo {author} {\bibfnamefont {L.}~\bibnamefont {Guo}},\ and\ \bibinfo {author} {\bibfnamefont {Y.}~\bibnamefont {Li}},\ }\href {https://doi.org/10.1103/PhysRevA.107.023705} {\bibfield  {journal} {\bibinfo  {journal} {Phys. Rev. A}\ }\textbf {\bibinfo {volume} {107}},\ \bibinfo {pages} {023705} (\bibinfo {year} {2023}{\natexlab{b}})}\BibitemShut {NoStop}%
\bibitem [{\citenamefont {Soro}\ and\ \citenamefont {Kockum}(2022)}]{DFI.Soro}%
  \BibitemOpen
  \bibfield  {author} {\bibinfo {author} {\bibfnamefont {A.}~\bibnamefont {Soro}}\ and\ \bibinfo {author} {\bibfnamefont {A.~F.}\ \bibnamefont {Kockum}},\ }\href {https://doi.org/10.1103/PhysRevA.105.023712} {\bibfield  {journal} {\bibinfo  {journal} {Phys. Rev. A}\ }\textbf {\bibinfo {volume} {105}},\ \bibinfo {pages} {023712} (\bibinfo {year} {2022})}\BibitemShut {NoStop}%
\bibitem [{\citenamefont {Soro}\ \emph {et~al.}(2023)\citenamefont {Soro}, \citenamefont {Mu\~noz},\ and\ \citenamefont {Kockum}}]{DFI.soro.2023}%
  \BibitemOpen
  \bibfield  {author} {\bibinfo {author} {\bibfnamefont {A.}~\bibnamefont {Soro}}, \bibinfo {author} {\bibfnamefont {C.~S.}\ \bibnamefont {Mu\~noz}},\ and\ \bibinfo {author} {\bibfnamefont {A.~F.}\ \bibnamefont {Kockum}},\ }\href {https://doi.org/10.1103/PhysRevA.107.013710} {\bibfield  {journal} {\bibinfo  {journal} {Phys. Rev. A}\ }\textbf {\bibinfo {volume} {107}},\ \bibinfo {pages} {013710} (\bibinfo {year} {2023})}\BibitemShut {NoStop}%
\bibitem [{\citenamefont {Andersson}\ \emph {et~al.}(2019{\natexlab{b}})\citenamefont {Andersson}, \citenamefont {Suri}, \citenamefont {Guo}, \citenamefont {Aref},\ and\ \citenamefont {Delsing}}]{NM.andersson}%
  \BibitemOpen
  \bibfield  {author} {\bibinfo {author} {\bibfnamefont {G.}~\bibnamefont {Andersson}}, \bibinfo {author} {\bibfnamefont {B.}~\bibnamefont {Suri}}, \bibinfo {author} {\bibfnamefont {L.}~\bibnamefont {Guo}}, \bibinfo {author} {\bibfnamefont {T.}~\bibnamefont {Aref}},\ and\ \bibinfo {author} {\bibfnamefont {P.}~\bibnamefont {Delsing}},\ }\href {https://doi.org/10.1038/s41567-019-0605-6} {\bibfield  {journal} {\bibinfo  {journal} {Nat. Phys.}\ }\textbf {\bibinfo {volume} {15}},\ \bibinfo {pages} {1123} (\bibinfo {year} {2019}{\natexlab{b}})}\BibitemShut {NoStop}%
\bibitem [{\citenamefont {Du}\ \emph {et~al.}(2022{\natexlab{a}})\citenamefont {Du}, \citenamefont {Chen}, \citenamefont {Zhang},\ and\ \citenamefont {Li}}]{NM.Du}%
  \BibitemOpen
  \bibfield  {author} {\bibinfo {author} {\bibfnamefont {L.}~\bibnamefont {Du}}, \bibinfo {author} {\bibfnamefont {Y.-T.}\ \bibnamefont {Chen}}, \bibinfo {author} {\bibfnamefont {Y.}~\bibnamefont {Zhang}},\ and\ \bibinfo {author} {\bibfnamefont {Y.}~\bibnamefont {Li}},\ }\href {https://doi.org/10.1103/PhysRevResearch.4.023198} {\bibfield  {journal} {\bibinfo  {journal} {Phys. Rev. Res.}\ }\textbf {\bibinfo {volume} {4}},\ \bibinfo {pages} {023198} (\bibinfo {year} {2022}{\natexlab{a}})}\BibitemShut {NoStop}%
\bibitem [{\citenamefont {Du}\ \emph {et~al.}(2022{\natexlab{b}})\citenamefont {Du}, \citenamefont {Zhang},\ and\ \citenamefont {Li}}]{NM.Du2023}%
  \BibitemOpen
  \bibfield  {author} {\bibinfo {author} {\bibfnamefont {L.}~\bibnamefont {Du}}, \bibinfo {author} {\bibfnamefont {Y.}~\bibnamefont {Zhang}},\ and\ \bibinfo {author} {\bibfnamefont {Y.}~\bibnamefont {Li}},\ }\href {https://doi.org/10.1007/s11467-022-1215-9} {\bibfield  {journal} {\bibinfo  {journal} {Frontiers of Physics}\ }\textbf {\bibinfo {volume} {18}},\ \bibinfo {pages} {12301} (\bibinfo {year} {2022}{\natexlab{b}})}\BibitemShut {NoStop}%
\bibitem [{\citenamefont {Yin}\ \emph {et~al.}(2022)\citenamefont {Yin}, \citenamefont {Luo},\ and\ \citenamefont {Liao}}]{NM.Yin}%
  \BibitemOpen
  \bibfield  {author} {\bibinfo {author} {\bibfnamefont {X.-L.}\ \bibnamefont {Yin}}, \bibinfo {author} {\bibfnamefont {W.-B.}\ \bibnamefont {Luo}},\ and\ \bibinfo {author} {\bibfnamefont {J.-Q.}\ \bibnamefont {Liao}},\ }\href {https://doi.org/10.1103/PhysRevA.106.063703} {\bibfield  {journal} {\bibinfo  {journal} {Phys. Rev. A}\ }\textbf {\bibinfo {volume} {106}},\ \bibinfo {pages} {063703} (\bibinfo {year} {2022})}\BibitemShut {NoStop}%
\bibitem [{\citenamefont {Wang}\ and\ \citenamefont {Li}(2022)}]{Chiral.Wang}%
  \BibitemOpen
  \bibfield  {author} {\bibinfo {author} {\bibfnamefont {X.}~\bibnamefont {Wang}}\ and\ \bibinfo {author} {\bibfnamefont {H.-R.}\ \bibnamefont {Li}},\ }\href {https://doi.org/10.1088/2058-9565/ac6a04} {\bibfield  {journal} {\bibinfo  {journal} {Quantum Sci. Technol.}\ }\textbf {\bibinfo {volume} {7}},\ \bibinfo {pages} {035007} (\bibinfo {year} {2022})}\BibitemShut {NoStop}%
\bibitem [{\citenamefont {Li}\ \emph {et~al.}(2024{\natexlab{b}})\citenamefont {Li}, \citenamefont {Zhang}, \citenamefont {Du}, \citenamefont {Li},\ and\ \citenamefont {Wu}}]{chiralWu2024}%
  \BibitemOpen
  \bibfield  {author} {\bibinfo {author} {\bibfnamefont {S.-Y.}\ \bibnamefont {Li}}, \bibinfo {author} {\bibfnamefont {Z.-Q.}\ \bibnamefont {Zhang}}, \bibinfo {author} {\bibfnamefont {L.}~\bibnamefont {Du}}, \bibinfo {author} {\bibfnamefont {Y.}~\bibnamefont {Li}},\ and\ \bibinfo {author} {\bibfnamefont {H.}~\bibnamefont {Wu}},\ }\href {https://doi.org/10.1103/PhysRevA.109.063703} {\bibfield  {journal} {\bibinfo  {journal} {Phys. Rev. A}\ }\textbf {\bibinfo {volume} {109}},\ \bibinfo {pages} {063703} (\bibinfo {year} {2024}{\natexlab{b}})}\BibitemShut {NoStop}%
\bibitem [{\citenamefont {Chen}\ \emph {et~al.}(2024)\citenamefont {Chen}, \citenamefont {Du}, \citenamefont {Wang}, \citenamefont {Artoni}, \citenamefont {La~Rocca},\ and\ \citenamefont {Wu}}]{Chen2024}%
  \BibitemOpen
  \bibfield  {author} {\bibinfo {author} {\bibfnamefont {Y.-T.}\ \bibnamefont {Chen}}, \bibinfo {author} {\bibfnamefont {L.}~\bibnamefont {Du}}, \bibinfo {author} {\bibfnamefont {Z.}~\bibnamefont {Wang}}, \bibinfo {author} {\bibfnamefont {M.}~\bibnamefont {Artoni}}, \bibinfo {author} {\bibfnamefont {G.~C.}\ \bibnamefont {La~Rocca}},\ and\ \bibinfo {author} {\bibfnamefont {J.-H.}\ \bibnamefont {Wu}},\ }\href {https://doi.org/10.1103/PhysRevA.109.063710} {\bibfield  {journal} {\bibinfo  {journal} {Phys. Rev. A}\ }\textbf {\bibinfo {volume} {109}},\ \bibinfo {pages} {063710} (\bibinfo {year} {2024})}\BibitemShut {NoStop}%
\bibitem [{\citenamefont {Zhao}\ and\ \citenamefont {Wang}(2020)}]{Zhao2020}%
  \BibitemOpen
  \bibfield  {author} {\bibinfo {author} {\bibfnamefont {W.}~\bibnamefont {Zhao}}\ and\ \bibinfo {author} {\bibfnamefont {Z.}~\bibnamefont {Wang}},\ }\href {https://doi.org/10.1103/PhysRevA.101.053855} {\bibfield  {journal} {\bibinfo  {journal} {Phys. Rev. A}\ }\textbf {\bibinfo {volume} {101}},\ \bibinfo {pages} {053855} (\bibinfo {year} {2020})}\BibitemShut {NoStop}%
\bibitem [{\citenamefont {Chen}\ \emph {et~al.}(2022)\citenamefont {Chen}, \citenamefont {Du}, \citenamefont {Guo}, \citenamefont {Wang}, \citenamefont {Zhang}, \citenamefont {Li},\ and\ \citenamefont {Wu}}]{chen2022}%
  \BibitemOpen
  \bibfield  {author} {\bibinfo {author} {\bibfnamefont {Y.-T.}\ \bibnamefont {Chen}}, \bibinfo {author} {\bibfnamefont {L.}~\bibnamefont {Du}}, \bibinfo {author} {\bibfnamefont {L.}~\bibnamefont {Guo}}, \bibinfo {author} {\bibfnamefont {Z.}~\bibnamefont {Wang}}, \bibinfo {author} {\bibfnamefont {Y.}~\bibnamefont {Zhang}}, \bibinfo {author} {\bibfnamefont {Y.}~\bibnamefont {Li}},\ and\ \bibinfo {author} {\bibfnamefont {J.-H.}\ \bibnamefont {Wu}},\ }\href {https://doi.org/10.1038/s42005-022-00991-3} {\bibfield  {journal} {\bibinfo  {journal} {Commun. Phys.}\ }\textbf {\bibinfo {volume} {5}},\ \bibinfo {pages} {215} (\bibinfo {year} {2022})}\BibitemShut {NoStop}%
\bibitem [{\citenamefont {Xu}\ and\ \citenamefont {Guo}(2025)}]{Xu2025}%
  \BibitemOpen
  \bibfield  {author} {\bibinfo {author} {\bibfnamefont {L.}~\bibnamefont {Xu}}\ and\ \bibinfo {author} {\bibfnamefont {L.}~\bibnamefont {Guo}},\ }\href {https://arxiv.org/abs/2502.08156} {\bibfield  {journal} {\bibinfo  {journal} {arXiv preprint arXiv:2502.08156}\ } (\bibinfo {year} {2025})}\BibitemShut {NoStop}%
\bibitem [{\citenamefont {Weng}\ \emph {et~al.}(2024)\citenamefont {Weng}, \citenamefont {Wang},\ and\ \citenamefont {Wang}}]{weng2024}%
  \BibitemOpen
  \bibfield  {author} {\bibinfo {author} {\bibfnamefont {M.}~\bibnamefont {Weng}}, \bibinfo {author} {\bibfnamefont {X.}~\bibnamefont {Wang}},\ and\ \bibinfo {author} {\bibfnamefont {Z.}~\bibnamefont {Wang}},\ }\href {https://doi.org/10.1103/PhysRevA.110.023721} {\bibfield  {journal} {\bibinfo  {journal} {Phys. Rev. A}\ }\textbf {\bibinfo {volume} {110}},\ \bibinfo {pages} {023721} (\bibinfo {year} {2024})}\BibitemShut {NoStop}%
\bibitem [{\citenamefont {Zhu}\ \emph {et~al.}(2024)\citenamefont {Zhu}, \citenamefont {Yin},\ and\ \citenamefont {Liao}}]{zhu2024}%
  \BibitemOpen
  \bibfield  {author} {\bibinfo {author} {\bibfnamefont {H.}~\bibnamefont {Zhu}}, \bibinfo {author} {\bibfnamefont {X.-L.}\ \bibnamefont {Yin}},\ and\ \bibinfo {author} {\bibfnamefont {J.-Q.}\ \bibnamefont {Liao}},\ }\href {https://arxiv.org/abs/2408.14178} {\bibfield  {journal} {\bibinfo  {journal} {arXiv preprint arXiv:2502.08156}\ } (\bibinfo {year} {2024})}\BibitemShut {NoStop}%
\bibitem [{\citenamefont {Wang}\ \emph {et~al.}(2024)\citenamefont {Wang}, \citenamefont {Zhu}, \citenamefont {Liu},\ and\ \citenamefont {Nori}}]{wang2024}%
  \BibitemOpen
  \bibfield  {author} {\bibinfo {author} {\bibfnamefont {X.}~\bibnamefont {Wang}}, \bibinfo {author} {\bibfnamefont {H.-B.}\ \bibnamefont {Zhu}}, \bibinfo {author} {\bibfnamefont {T.}~\bibnamefont {Liu}},\ and\ \bibinfo {author} {\bibfnamefont {F.}~\bibnamefont {Nori}},\ }\href {https://doi.org/10.1103/PhysRevResearch.6.013279} {\bibfield  {journal} {\bibinfo  {journal} {Phys. Rev. Res.}\ }\textbf {\bibinfo {volume} {6}},\ \bibinfo {pages} {013279} (\bibinfo {year} {2024})}\BibitemShut {NoStop}%
\bibitem [{\citenamefont {Yuan}\ \emph {et~al.}(2018{\natexlab{a}})\citenamefont {Yuan}, \citenamefont {Lin}, \citenamefont {Xiao},\ and\ \citenamefont {Fan}}]{SD.Yuan}%
  \BibitemOpen
  \bibfield  {author} {\bibinfo {author} {\bibfnamefont {L.}~\bibnamefont {Yuan}}, \bibinfo {author} {\bibfnamefont {Q.}~\bibnamefont {Lin}}, \bibinfo {author} {\bibfnamefont {M.}~\bibnamefont {Xiao}},\ and\ \bibinfo {author} {\bibfnamefont {S.}~\bibnamefont {Fan}},\ }\href {https://doi.org/10.1364/OPTICA.5.001396} {\bibfield  {journal} {\bibinfo  {journal} {Optica}\ }\textbf {\bibinfo {volume} {5}},\ \bibinfo {pages} {1396} (\bibinfo {year} {2018}{\natexlab{a}})}\BibitemShut {NoStop}%
\bibitem [{\citenamefont {Celi}\ \emph {et~al.}(2014)\citenamefont {Celi}, \citenamefont {Massignan}, \citenamefont {Ruseckas}, \citenamefont {Goldman}, \citenamefont {Spielman}, \citenamefont {Juzeli\ifmmode~\bar{u}\else \={u}\fi{}nas},\ and\ \citenamefont {Lewenstein}}]{ColdAtom.Celi}%
  \BibitemOpen
  \bibfield  {author} {\bibinfo {author} {\bibfnamefont {A.}~\bibnamefont {Celi}}, \bibinfo {author} {\bibfnamefont {P.}~\bibnamefont {Massignan}}, \bibinfo {author} {\bibfnamefont {J.}~\bibnamefont {Ruseckas}}, \bibinfo {author} {\bibfnamefont {N.}~\bibnamefont {Goldman}}, \bibinfo {author} {\bibfnamefont {I.~B.}\ \bibnamefont {Spielman}}, \bibinfo {author} {\bibfnamefont {G.}~\bibnamefont {Juzeli\ifmmode~\bar{u}\else \={u}\fi{}nas}},\ and\ \bibinfo {author} {\bibfnamefont {M.}~\bibnamefont {Lewenstein}},\ }\href {https://doi.org/10.1103/PhysRevLett.112.043001} {\bibfield  {journal} {\bibinfo  {journal} {Phys. Rev. Lett.}\ }\textbf {\bibinfo {volume} {112}},\ \bibinfo {pages} {043001} (\bibinfo {year} {2014})}\BibitemShut {NoStop}%
\bibitem [{\citenamefont {Price}\ \emph {et~al.}(2017)\citenamefont {Price}, \citenamefont {Ozawa},\ and\ \citenamefont {Goldman}}]{ColdAtom.Price}%
  \BibitemOpen
  \bibfield  {author} {\bibinfo {author} {\bibfnamefont {H.~M.}\ \bibnamefont {Price}}, \bibinfo {author} {\bibfnamefont {T.}~\bibnamefont {Ozawa}},\ and\ \bibinfo {author} {\bibfnamefont {N.}~\bibnamefont {Goldman}},\ }\href {https://doi.org/10.1103/PhysRevA.95.023607} {\bibfield  {journal} {\bibinfo  {journal} {Phys. Rev. A}\ }\textbf {\bibinfo {volume} {95}},\ \bibinfo {pages} {023607} (\bibinfo {year} {2017})}\BibitemShut {NoStop}%
\bibitem [{\citenamefont {Price}\ \emph {et~al.}(2015)\citenamefont {Price}, \citenamefont {Zilberberg}, \citenamefont {Ozawa}, \citenamefont {Carusotto},\ and\ \citenamefont {Goldman}}]{ColdAtom.Price2015}%
  \BibitemOpen
  \bibfield  {author} {\bibinfo {author} {\bibfnamefont {H.~M.}\ \bibnamefont {Price}}, \bibinfo {author} {\bibfnamefont {O.}~\bibnamefont {Zilberberg}}, \bibinfo {author} {\bibfnamefont {T.}~\bibnamefont {Ozawa}}, \bibinfo {author} {\bibfnamefont {I.}~\bibnamefont {Carusotto}},\ and\ \bibinfo {author} {\bibfnamefont {N.}~\bibnamefont {Goldman}},\ }\href {https://doi.org/10.1103/PhysRevLett.115.195303} {\bibfield  {journal} {\bibinfo  {journal} {Phys. Rev. Lett.}\ }\textbf {\bibinfo {volume} {115}},\ \bibinfo {pages} {195303} (\bibinfo {year} {2015})}\BibitemShut {NoStop}%
\bibitem [{\citenamefont {Mei}\ \emph {et~al.}(2016)\citenamefont {Mei}, \citenamefont {Xue}, \citenamefont {Zhang}, \citenamefont {Tian}, \citenamefont {Lee},\ and\ \citenamefont {Zhu}}]{SQC.Feng}%
  \BibitemOpen
  \bibfield  {author} {\bibinfo {author} {\bibfnamefont {F.}~\bibnamefont {Mei}}, \bibinfo {author} {\bibfnamefont {Z.-Y.}\ \bibnamefont {Xue}}, \bibinfo {author} {\bibfnamefont {D.-W.}\ \bibnamefont {Zhang}}, \bibinfo {author} {\bibfnamefont {L.}~\bibnamefont {Tian}}, \bibinfo {author} {\bibfnamefont {C.}~\bibnamefont {Lee}},\ and\ \bibinfo {author} {\bibfnamefont {S.-L.}\ \bibnamefont {Zhu}},\ }\href {https://doi.org/10.1088/2058-9565/1/1/015006} {\bibfield  {journal} {\bibinfo  {journal} {Quantum Sci. Technol.}\ }\textbf {\bibinfo {volume} {1}},\ \bibinfo {pages} {015006} (\bibinfo {year} {2016})}\BibitemShut {NoStop}%
\bibitem [{\citenamefont {Tsomokos}\ \emph {et~al.}(2010)\citenamefont {Tsomokos}, \citenamefont {Ashhab},\ and\ \citenamefont {Nori}}]{SQC.Tsomokos}%
  \BibitemOpen
  \bibfield  {author} {\bibinfo {author} {\bibfnamefont {D.~I.}\ \bibnamefont {Tsomokos}}, \bibinfo {author} {\bibfnamefont {S.}~\bibnamefont {Ashhab}},\ and\ \bibinfo {author} {\bibfnamefont {F.}~\bibnamefont {Nori}},\ }\href {https://doi.org/10.1103/PhysRevA.82.052311} {\bibfield  {journal} {\bibinfo  {journal} {Phys. Rev. A}\ }\textbf {\bibinfo {volume} {82}},\ \bibinfo {pages} {052311} (\bibinfo {year} {2010})}\BibitemShut {NoStop}%
\bibitem [{\citenamefont {Zhou}\ \emph {et~al.}(2017)\citenamefont {Zhou}, \citenamefont {Luo}, \citenamefont {Wang}, \citenamefont {Guo}, \citenamefont {Zhou}, \citenamefont {Pu},\ and\ \citenamefont {Zhou}}]{PS.Zhou}%
  \BibitemOpen
  \bibfield  {author} {\bibinfo {author} {\bibfnamefont {X.-F.}\ \bibnamefont {Zhou}}, \bibinfo {author} {\bibfnamefont {X.-W.}\ \bibnamefont {Luo}}, \bibinfo {author} {\bibfnamefont {S.}~\bibnamefont {Wang}}, \bibinfo {author} {\bibfnamefont {G.-C.}\ \bibnamefont {Guo}}, \bibinfo {author} {\bibfnamefont {X.}~\bibnamefont {Zhou}}, \bibinfo {author} {\bibfnamefont {H.}~\bibnamefont {Pu}},\ and\ \bibinfo {author} {\bibfnamefont {Z.-W.}\ \bibnamefont {Zhou}},\ }\href {https://doi.org/10.1103/PhysRevLett.118.083603} {\bibfield  {journal} {\bibinfo  {journal} {Phys. Rev. Lett.}\ }\textbf {\bibinfo {volume} {118}},\ \bibinfo {pages} {083603} (\bibinfo {year} {2017})}\BibitemShut {NoStop}%
\bibitem [{\citenamefont {Ozawa}\ \emph {et~al.}(2016)\citenamefont {Ozawa}, \citenamefont {Price}, \citenamefont {Goldman}, \citenamefont {Zilberberg},\ and\ \citenamefont {Carusotto}}]{PS.Ozawa}%
  \BibitemOpen
  \bibfield  {author} {\bibinfo {author} {\bibfnamefont {T.}~\bibnamefont {Ozawa}}, \bibinfo {author} {\bibfnamefont {H.~M.}\ \bibnamefont {Price}}, \bibinfo {author} {\bibfnamefont {N.}~\bibnamefont {Goldman}}, \bibinfo {author} {\bibfnamefont {O.}~\bibnamefont {Zilberberg}},\ and\ \bibinfo {author} {\bibfnamefont {I.}~\bibnamefont {Carusotto}},\ }\href {https://doi.org/10.1103/PhysRevA.93.043827} {\bibfield  {journal} {\bibinfo  {journal} {Phys. Rev. A}\ }\textbf {\bibinfo {volume} {93}},\ \bibinfo {pages} {043827} (\bibinfo {year} {2016})}\BibitemShut {NoStop}%
\bibitem [{\citenamefont {Balčytis}\ \emph {et~al.}(2022)\citenamefont {Balčytis}, \citenamefont {Ozawa}, \citenamefont {Ota}, \citenamefont {Iwamoto}, \citenamefont {Maeda},\ and\ \citenamefont {Baba}}]{PS.Armandas}%
  \BibitemOpen
  \bibfield  {author} {\bibinfo {author} {\bibfnamefont {A.}~\bibnamefont {Balčytis}}, \bibinfo {author} {\bibfnamefont {T.}~\bibnamefont {Ozawa}}, \bibinfo {author} {\bibfnamefont {Y.}~\bibnamefont {Ota}}, \bibinfo {author} {\bibfnamefont {S.}~\bibnamefont {Iwamoto}}, \bibinfo {author} {\bibfnamefont {J.}~\bibnamefont {Maeda}},\ and\ \bibinfo {author} {\bibfnamefont {T.}~\bibnamefont {Baba}},\ }\href {https://doi.org/10.1126/sciadv.abk0468} {\bibfield  {journal} {\bibinfo  {journal} {Sci. Adv.}\ }\textbf {\bibinfo {volume} {8}},\ \bibinfo {pages} {eabk0468} (\bibinfo {year} {2022})}\BibitemShut {NoStop}%
\bibitem [{\citenamefont {Yuan}\ and\ \citenamefont {Fan}(2016)}]{FD.Yuan}%
  \BibitemOpen
  \bibfield  {author} {\bibinfo {author} {\bibfnamefont {L.}~\bibnamefont {Yuan}}\ and\ \bibinfo {author} {\bibfnamefont {S.}~\bibnamefont {Fan}},\ }\href {https://doi.org/10.1364/OPTICA.3.001014} {\bibfield  {journal} {\bibinfo  {journal} {Optica}\ }\textbf {\bibinfo {volume} {3}},\ \bibinfo {pages} {1014} (\bibinfo {year} {2016})}\BibitemShut {NoStop}%
\bibitem [{\citenamefont {Qin}\ \emph {et~al.}(2018)\citenamefont {Qin}, \citenamefont {Zhou}, \citenamefont {Peng}, \citenamefont {Sounas}, \citenamefont {Zhu}, \citenamefont {Wang}, \citenamefont {Dong}, \citenamefont {Zhang}, \citenamefont {Al\`u},\ and\ \citenamefont {Lu}}]{FD.Qin}%
  \BibitemOpen
  \bibfield  {author} {\bibinfo {author} {\bibfnamefont {C.}~\bibnamefont {Qin}}, \bibinfo {author} {\bibfnamefont {F.}~\bibnamefont {Zhou}}, \bibinfo {author} {\bibfnamefont {Y.}~\bibnamefont {Peng}}, \bibinfo {author} {\bibfnamefont {D.}~\bibnamefont {Sounas}}, \bibinfo {author} {\bibfnamefont {X.}~\bibnamefont {Zhu}}, \bibinfo {author} {\bibfnamefont {B.}~\bibnamefont {Wang}}, \bibinfo {author} {\bibfnamefont {J.}~\bibnamefont {Dong}}, \bibinfo {author} {\bibfnamefont {X.}~\bibnamefont {Zhang}}, \bibinfo {author} {\bibfnamefont {A.}~\bibnamefont {Al\`u}},\ and\ \bibinfo {author} {\bibfnamefont {P.}~\bibnamefont {Lu}},\ }\href {https://doi.org/10.1103/PhysRevLett.120.133901} {\bibfield  {journal} {\bibinfo  {journal} {Phys. Rev. Lett.}\ }\textbf {\bibinfo {volume} {120}},\ \bibinfo {pages} {133901} (\bibinfo {year} {2018})}\BibitemShut {NoStop}%
\bibitem [{\citenamefont {Lin}\ \emph {et~al.}(2018)\citenamefont {Lin}, \citenamefont {Sun}, \citenamefont {Xiao}, \citenamefont {Zhang},\ and\ \citenamefont {Fan}}]{FD.Qian}%
  \BibitemOpen
  \bibfield  {author} {\bibinfo {author} {\bibfnamefont {Q.}~\bibnamefont {Lin}}, \bibinfo {author} {\bibfnamefont {X.-Q.}\ \bibnamefont {Sun}}, \bibinfo {author} {\bibfnamefont {M.}~\bibnamefont {Xiao}}, \bibinfo {author} {\bibfnamefont {S.-C.}\ \bibnamefont {Zhang}},\ and\ \bibinfo {author} {\bibfnamefont {S.}~\bibnamefont {Fan}},\ }\href {https://doi.org/10.1126/sciadv.aat2774} {\bibfield  {journal} {\bibinfo  {journal} {Sci. Adv.}\ }\textbf {\bibinfo {volume} {4}},\ \bibinfo {pages} {eaat2774} (\bibinfo {year} {2018})}\BibitemShut {NoStop}%
\bibitem [{\citenamefont {Yuan}\ \emph {et~al.}(2018{\natexlab{b}})\citenamefont {Yuan}, \citenamefont {Lin}, \citenamefont {Xiao}, \citenamefont {Dutt},\ and\ \citenamefont {Fan}}]{FD.Yuan18}%
  \BibitemOpen
  \bibfield  {author} {\bibinfo {author} {\bibfnamefont {L.}~\bibnamefont {Yuan}}, \bibinfo {author} {\bibfnamefont {Q.}~\bibnamefont {Lin}}, \bibinfo {author} {\bibfnamefont {M.}~\bibnamefont {Xiao}}, \bibinfo {author} {\bibfnamefont {A.}~\bibnamefont {Dutt}},\ and\ \bibinfo {author} {\bibfnamefont {S.}~\bibnamefont {Fan}},\ }\href {https://doi.org/10.1063/1.5039375} {\bibfield  {journal} {\bibinfo  {journal} {APL Photonics}\ }\textbf {\bibinfo {volume} {3}},\ \bibinfo {pages} {086103} (\bibinfo {year} {2018}{\natexlab{b}})}\BibitemShut {NoStop}%
\bibitem [{\citenamefont {Yuan}\ \emph {et~al.}(2018{\natexlab{c}})\citenamefont {Yuan}, \citenamefont {Xiao}, \citenamefont {Lin},\ and\ \citenamefont {Fan}}]{FD.Yuan2018}%
  \BibitemOpen
  \bibfield  {author} {\bibinfo {author} {\bibfnamefont {L.}~\bibnamefont {Yuan}}, \bibinfo {author} {\bibfnamefont {M.}~\bibnamefont {Xiao}}, \bibinfo {author} {\bibfnamefont {Q.}~\bibnamefont {Lin}},\ and\ \bibinfo {author} {\bibfnamefont {S.}~\bibnamefont {Fan}},\ }\href {https://doi.org/10.1103/PhysRevB.97.104105} {\bibfield  {journal} {\bibinfo  {journal} {Phys. Rev. B}\ }\textbf {\bibinfo {volume} {97}},\ \bibinfo {pages} {104105} (\bibinfo {year} {2018}{\natexlab{c}})}\BibitemShut {NoStop}%
\bibitem [{\citenamefont {Yuan}\ \emph {et~al.}(2021)\citenamefont {Yuan}, \citenamefont {Dutt},\ and\ \citenamefont {Fan}}]{FD.yuan2021}%
  \BibitemOpen
  \bibfield  {author} {\bibinfo {author} {\bibfnamefont {L.}~\bibnamefont {Yuan}}, \bibinfo {author} {\bibfnamefont {A.}~\bibnamefont {Dutt}},\ and\ \bibinfo {author} {\bibfnamefont {S.}~\bibnamefont {Fan}},\ }\href {https://pubs.aip.org/aip/app/article/6/7/071102/892590} {\bibfield  {journal} {\bibinfo  {journal} {APL Photonics}\ }\textbf {\bibinfo {volume} {6}} (\bibinfo {year} {2021})}\BibitemShut {NoStop}%
\bibitem [{\citenamefont {Yuan}\ \emph {et~al.}(2019)\citenamefont {Yuan}, \citenamefont {Lin}, \citenamefont {Zhang}, \citenamefont {Xiao}, \citenamefont {Chen},\ and\ \citenamefont {Fan}}]{OAM.Yuan}%
  \BibitemOpen
  \bibfield  {author} {\bibinfo {author} {\bibfnamefont {L.}~\bibnamefont {Yuan}}, \bibinfo {author} {\bibfnamefont {Q.}~\bibnamefont {Lin}}, \bibinfo {author} {\bibfnamefont {A.}~\bibnamefont {Zhang}}, \bibinfo {author} {\bibfnamefont {M.}~\bibnamefont {Xiao}}, \bibinfo {author} {\bibfnamefont {X.}~\bibnamefont {Chen}},\ and\ \bibinfo {author} {\bibfnamefont {S.}~\bibnamefont {Fan}},\ }\href {https://doi.org/10.1103/PhysRevLett.122.083903} {\bibfield  {journal} {\bibinfo  {journal} {Phys. Rev. Lett.}\ }\textbf {\bibinfo {volume} {122}},\ \bibinfo {pages} {083903} (\bibinfo {year} {2019})}\BibitemShut {NoStop}%
\bibitem [{\citenamefont {Lustig}\ \emph {et~al.}(2019)\citenamefont {Lustig}, \citenamefont {Weimann}, \citenamefont {Plotnik}, \citenamefont {Lumer}, \citenamefont {Bandres}, \citenamefont {Szameit},\ and\ \citenamefont {Segev}}]{PS.lustig}%
  \BibitemOpen
  \bibfield  {author} {\bibinfo {author} {\bibfnamefont {E.}~\bibnamefont {Lustig}}, \bibinfo {author} {\bibfnamefont {S.}~\bibnamefont {Weimann}}, \bibinfo {author} {\bibfnamefont {Y.}~\bibnamefont {Plotnik}}, \bibinfo {author} {\bibfnamefont {Y.}~\bibnamefont {Lumer}}, \bibinfo {author} {\bibfnamefont {M.~A.}\ \bibnamefont {Bandres}}, \bibinfo {author} {\bibfnamefont {A.}~\bibnamefont {Szameit}},\ and\ \bibinfo {author} {\bibfnamefont {M.}~\bibnamefont {Segev}},\ }\href {https://doi.org/10.1038/s41586-019-0943-7} {\bibfield  {journal} {\bibinfo  {journal} {Nature (London)}\ }\textbf {\bibinfo {volume} {567}},\ \bibinfo {pages} {356} (\bibinfo {year} {2019})}\BibitemShut {NoStop}%
\bibitem [{\citenamefont {Luo}\ \emph {et~al.}(2017)\citenamefont {Luo}, \citenamefont {Zhou}, \citenamefont {Xu}, \citenamefont {Li}, \citenamefont {Guo}, \citenamefont {Zhang},\ and\ \citenamefont {Zhou}}]{PS.luo}%
  \BibitemOpen
  \bibfield  {author} {\bibinfo {author} {\bibfnamefont {X.-W.}\ \bibnamefont {Luo}}, \bibinfo {author} {\bibfnamefont {X.}~\bibnamefont {Zhou}}, \bibinfo {author} {\bibfnamefont {J.-S.}\ \bibnamefont {Xu}}, \bibinfo {author} {\bibfnamefont {C.-F.}\ \bibnamefont {Li}}, \bibinfo {author} {\bibfnamefont {G.-C.}\ \bibnamefont {Guo}}, \bibinfo {author} {\bibfnamefont {C.}~\bibnamefont {Zhang}},\ and\ \bibinfo {author} {\bibfnamefont {Z.-W.}\ \bibnamefont {Zhou}},\ }\href {https://doi.org/10.1038/ncomms16097} {\bibfield  {journal} {\bibinfo  {journal} {Nat. Commun.}\ }\textbf {\bibinfo {volume} {8}},\ \bibinfo {pages} {16097} (\bibinfo {year} {2017})}\BibitemShut {NoStop}%
\bibitem [{\citenamefont {Yang}\ \emph {et~al.}(2023)\citenamefont {Yang}, \citenamefont {Zhang}, \citenamefont {Liao}, \citenamefont {Liu}, \citenamefont {Zhou}, \citenamefont {Zhou}, \citenamefont {Xu}, \citenamefont {Han}, \citenamefont {Li},\ and\ \citenamefont {Guo}}]{OAM.Mu}%
  \BibitemOpen
  \bibfield  {author} {\bibinfo {author} {\bibfnamefont {M.}~\bibnamefont {Yang}}, \bibinfo {author} {\bibfnamefont {H.-Q.}\ \bibnamefont {Zhang}}, \bibinfo {author} {\bibfnamefont {Y.-W.}\ \bibnamefont {Liao}}, \bibinfo {author} {\bibfnamefont {Z.-H.}\ \bibnamefont {Liu}}, \bibinfo {author} {\bibfnamefont {Z.-W.}\ \bibnamefont {Zhou}}, \bibinfo {author} {\bibfnamefont {X.-X.}\ \bibnamefont {Zhou}}, \bibinfo {author} {\bibfnamefont {J.-S.}\ \bibnamefont {Xu}}, \bibinfo {author} {\bibfnamefont {Y.-J.}\ \bibnamefont {Han}}, \bibinfo {author} {\bibfnamefont {C.-F.}\ \bibnamefont {Li}},\ and\ \bibinfo {author} {\bibfnamefont {G.-C.}\ \bibnamefont {Guo}},\ }\href {https://doi.org/10.1126/sciadv.abp8943} {\bibfield  {journal} {\bibinfo  {journal} {Sci. Adv.}\ }\textbf {\bibinfo {volume} {9}},\ \bibinfo {pages} {eabp8943} (\bibinfo {year} {2023})}\BibitemShut {NoStop}%
\bibitem [{\citenamefont {Yuan}\ \emph {et~al.}(2016)\citenamefont {Yuan}, \citenamefont {Shi},\ and\ \citenamefont {Fan}}]{MP.Yuan}%
  \BibitemOpen
  \bibfield  {author} {\bibinfo {author} {\bibfnamefont {L.}~\bibnamefont {Yuan}}, \bibinfo {author} {\bibfnamefont {Y.}~\bibnamefont {Shi}},\ and\ \bibinfo {author} {\bibfnamefont {S.}~\bibnamefont {Fan}},\ }\href {https://doi.org/10.1364/OL.41.000741} {\bibfield  {journal} {\bibinfo  {journal} {Opt. Lett.}\ }\textbf {\bibinfo {volume} {41}},\ \bibinfo {pages} {741} (\bibinfo {year} {2016})}\BibitemShut {NoStop}%
\bibitem [{\citenamefont {Regensburger}\ \emph {et~al.}(2013)\citenamefont {Regensburger}, \citenamefont {Miri}, \citenamefont {Bersch}, \citenamefont {N\"ager}, \citenamefont {Onishchukov}, \citenamefont {Christodoulides},\ and\ \citenamefont {Peschel}}]{MP_Alois}%
  \BibitemOpen
  \bibfield  {author} {\bibinfo {author} {\bibfnamefont {A.}~\bibnamefont {Regensburger}}, \bibinfo {author} {\bibfnamefont {M.-A.}\ \bibnamefont {Miri}}, \bibinfo {author} {\bibfnamefont {C.}~\bibnamefont {Bersch}}, \bibinfo {author} {\bibfnamefont {J.}~\bibnamefont {N\"ager}}, \bibinfo {author} {\bibfnamefont {G.}~\bibnamefont {Onishchukov}}, \bibinfo {author} {\bibfnamefont {D.~N.}\ \bibnamefont {Christodoulides}},\ and\ \bibinfo {author} {\bibfnamefont {U.}~\bibnamefont {Peschel}},\ }\href {https://doi.org/10.1103/PhysRevLett.110.223902} {\bibfield  {journal} {\bibinfo  {journal} {Phys. Rev. Lett.}\ }\textbf {\bibinfo {volume} {110}},\ \bibinfo {pages} {223902} (\bibinfo {year} {2013})}\BibitemShut {NoStop}%
\bibitem [{\citenamefont {Wimmer}\ \emph {et~al.}(2013)\citenamefont {Wimmer}, \citenamefont {Regensburger}, \citenamefont {Bersch}, \citenamefont {Miri}, \citenamefont {Batz}, \citenamefont {Onishchukov}, \citenamefont {Christodoulides},\ and\ \citenamefont {Peschel}}]{MP_wimmer2013}%
  \BibitemOpen
  \bibfield  {author} {\bibinfo {author} {\bibfnamefont {M.}~\bibnamefont {Wimmer}}, \bibinfo {author} {\bibfnamefont {A.}~\bibnamefont {Regensburger}}, \bibinfo {author} {\bibfnamefont {C.}~\bibnamefont {Bersch}}, \bibinfo {author} {\bibfnamefont {M.-A.}\ \bibnamefont {Miri}}, \bibinfo {author} {\bibfnamefont {S.}~\bibnamefont {Batz}}, \bibinfo {author} {\bibfnamefont {G.}~\bibnamefont {Onishchukov}}, \bibinfo {author} {\bibfnamefont {D.~N.}\ \bibnamefont {Christodoulides}},\ and\ \bibinfo {author} {\bibfnamefont {U.}~\bibnamefont {Peschel}},\ }\href {https://doi.org/10.1038/nphys2777} {\bibfield  {journal} {\bibinfo  {journal} {Nat. Phys.}\ }\textbf {\bibinfo {volume} {9}},\ \bibinfo {pages} {780} (\bibinfo {year} {2013})}\BibitemShut {NoStop}%
\bibitem [{\citenamefont {Wimmer}\ \emph {et~al.}(2015)\citenamefont {Wimmer}, \citenamefont {Regensburger}, \citenamefont {Miri}, \citenamefont {Bersch}, \citenamefont {Christodoulides},\ and\ \citenamefont {Peschel}}]{MP_wimmer2015}%
  \BibitemOpen
  \bibfield  {author} {\bibinfo {author} {\bibfnamefont {M.}~\bibnamefont {Wimmer}}, \bibinfo {author} {\bibfnamefont {A.}~\bibnamefont {Regensburger}}, \bibinfo {author} {\bibfnamefont {M.-A.}\ \bibnamefont {Miri}}, \bibinfo {author} {\bibfnamefont {C.}~\bibnamefont {Bersch}}, \bibinfo {author} {\bibfnamefont {D.~N.}\ \bibnamefont {Christodoulides}},\ and\ \bibinfo {author} {\bibfnamefont {U.}~\bibnamefont {Peschel}},\ }\href {https://doi.org/10.1038/ncomms8782} {\bibfield  {journal} {\bibinfo  {journal} {Nat. Commun.}\ }\textbf {\bibinfo {volume} {6}},\ \bibinfo {pages} {7782} (\bibinfo {year} {2015})}\BibitemShut {NoStop}%
\bibitem [{\citenamefont {Du}\ \emph {et~al.}(2022{\natexlab{c}})\citenamefont {Du}, \citenamefont {Zhang}, \citenamefont {Wu}, \citenamefont {Kockum},\ and\ \citenamefont {Li}}]{GiantAtom.Lei}%
  \BibitemOpen
  \bibfield  {author} {\bibinfo {author} {\bibfnamefont {L.}~\bibnamefont {Du}}, \bibinfo {author} {\bibfnamefont {Y.}~\bibnamefont {Zhang}}, \bibinfo {author} {\bibfnamefont {J.-H.}\ \bibnamefont {Wu}}, \bibinfo {author} {\bibfnamefont {A.~F.}\ \bibnamefont {Kockum}},\ and\ \bibinfo {author} {\bibfnamefont {Y.}~\bibnamefont {Li}},\ }\href {https://doi.org/10.1103/PhysRevLett.128.223602} {\bibfield  {journal} {\bibinfo  {journal} {Phys. Rev. Lett.}\ }\textbf {\bibinfo {volume} {128}},\ \bibinfo {pages} {223602} (\bibinfo {year} {2022}{\natexlab{c}})}\BibitemShut {NoStop}%
\bibitem [{\citenamefont {Murali}\ \emph {et~al.}(2004)\citenamefont {Murali}, \citenamefont {Dutton}, \citenamefont {Oliver}, \citenamefont {Crankshaw},\ and\ \citenamefont {Orlando}}]{SQC.Murali}%
  \BibitemOpen
  \bibfield  {author} {\bibinfo {author} {\bibfnamefont {K.~V. R.~M.}\ \bibnamefont {Murali}}, \bibinfo {author} {\bibfnamefont {Z.}~\bibnamefont {Dutton}}, \bibinfo {author} {\bibfnamefont {W.~D.}\ \bibnamefont {Oliver}}, \bibinfo {author} {\bibfnamefont {D.~S.}\ \bibnamefont {Crankshaw}},\ and\ \bibinfo {author} {\bibfnamefont {T.~P.}\ \bibnamefont {Orlando}},\ }\href {https://doi.org/10.1103/PhysRevLett.93.087003} {\bibfield  {journal} {\bibinfo  {journal} {Phys. Rev. Lett.}\ }\textbf {\bibinfo {volume} {93}},\ \bibinfo {pages} {087003} (\bibinfo {year} {2004})}\BibitemShut {NoStop}%
\bibitem [{\citenamefont {Blais}\ \emph {et~al.}(2021)\citenamefont {Blais}, \citenamefont {Grimsmo}, \citenamefont {Girvin},\ and\ \citenamefont {Wallraff}}]{SQC.Wallraff}%
  \BibitemOpen
  \bibfield  {author} {\bibinfo {author} {\bibfnamefont {A.}~\bibnamefont {Blais}}, \bibinfo {author} {\bibfnamefont {A.~L.}\ \bibnamefont {Grimsmo}}, \bibinfo {author} {\bibfnamefont {S.~M.}\ \bibnamefont {Girvin}},\ and\ \bibinfo {author} {\bibfnamefont {A.}~\bibnamefont {Wallraff}},\ }\href {https://doi.org/10.1103/RevModPhys.93.025005} {\bibfield  {journal} {\bibinfo  {journal} {Rev. Mod. Phys.}\ }\textbf {\bibinfo {volume} {93}},\ \bibinfo {pages} {025005} (\bibinfo {year} {2021})}\BibitemShut {NoStop}%
\bibitem [{\citenamefont {Zhang}\ \emph {et~al.}(2019)\citenamefont {Zhang}, \citenamefont {Buscaino}, \citenamefont {Wang}, \citenamefont {Shams-Ansari}, \citenamefont {Reimer}, \citenamefont {Zhu}, \citenamefont {Kahn},\ and\ \citenamefont {Lončar}}]{Broadband2019}%
  \BibitemOpen
  \bibfield  {author} {\bibinfo {author} {\bibfnamefont {M.}~\bibnamefont {Zhang}}, \bibinfo {author} {\bibfnamefont {B.}~\bibnamefont {Buscaino}}, \bibinfo {author} {\bibfnamefont {C.}~\bibnamefont {Wang}}, \bibinfo {author} {\bibfnamefont {A.}~\bibnamefont {Shams-Ansari}}, \bibinfo {author} {\bibfnamefont {C.}~\bibnamefont {Reimer}}, \bibinfo {author} {\bibfnamefont {R.}~\bibnamefont {Zhu}}, \bibinfo {author} {\bibfnamefont {J.~M.}\ \bibnamefont {Kahn}},\ and\ \bibinfo {author} {\bibfnamefont {M.}~\bibnamefont {Lončar}},\ }\href {https://doi.org/10.1038/s41586-019-1008-7} {\bibfield  {journal} {\bibinfo  {journal} {Nature (London)}\ }\textbf {\bibinfo {volume} {568}},\ \bibinfo {pages} {373} (\bibinfo {year} {2019})}\BibitemShut {NoStop}%
\bibitem [{\citenamefont {Zhao}\ \emph {et~al.}(2024)\citenamefont {Zhao}, \citenamefont {Jang}, \citenamefont {Beals}, \citenamefont {McNulty}, \citenamefont {Ji}, \citenamefont {Okawachi}, \citenamefont {Lipson},\ and\ \citenamefont {Gaeta}}]{frequencycomb}%
  \BibitemOpen
  \bibfield  {author} {\bibinfo {author} {\bibfnamefont {Y.}~\bibnamefont {Zhao}}, \bibinfo {author} {\bibfnamefont {J.~K.}\ \bibnamefont {Jang}}, \bibinfo {author} {\bibfnamefont {G.~J.}\ \bibnamefont {Beals}}, \bibinfo {author} {\bibfnamefont {K.~J.}\ \bibnamefont {McNulty}}, \bibinfo {author} {\bibfnamefont {X.}~\bibnamefont {Ji}}, \bibinfo {author} {\bibfnamefont {Y.}~\bibnamefont {Okawachi}}, \bibinfo {author} {\bibfnamefont {M.}~\bibnamefont {Lipson}},\ and\ \bibinfo {author} {\bibfnamefont {A.~L.}\ \bibnamefont {Gaeta}},\ }\href {https://www.nature.com/articles/s41586-024-07136-2} {\bibfield  {journal} {\bibinfo  {journal} {Nature (London)}\ }\textbf {\bibinfo {volume} {627}},\ \bibinfo {pages} {546} (\bibinfo {year} {2024})}\BibitemShut {NoStop}%
\bibitem [{\citenamefont {Xiao}\ \emph {et~al.}(2022)\citenamefont {Xiao}, \citenamefont {Wang}, \citenamefont {Li}, \citenamefont {Chen},\ and\ \citenamefont {Yuan}}]{xiao2022bound}%
  \BibitemOpen
  \bibfield  {author} {\bibinfo {author} {\bibfnamefont {H.}~\bibnamefont {Xiao}}, \bibinfo {author} {\bibfnamefont {L.}~\bibnamefont {Wang}}, \bibinfo {author} {\bibfnamefont {Z.-H.}\ \bibnamefont {Li}}, \bibinfo {author} {\bibfnamefont {X.}~\bibnamefont {Chen}},\ and\ \bibinfo {author} {\bibfnamefont {L.}~\bibnamefont {Yuan}},\ }\href {https://doi.org/10.1038/s41534-022-00591-7} {\bibfield  {journal} {\bibinfo  {journal} {npj Quantum Inf.}\ }\textbf {\bibinfo {volume} {8}},\ \bibinfo {pages} {80} (\bibinfo {year} {2022})}\BibitemShut {NoStop}%
\bibitem [{\citenamefont {Gong}\ \emph {et~al.}(2024)\citenamefont {Gong}, \citenamefont {He}, \citenamefont {Yu}, \citenamefont {Zhang}, \citenamefont {Nori},\ and\ \citenamefont {Xiang}}]{gong2024}%
  \BibitemOpen
  \bibfield  {author} {\bibinfo {author} {\bibfnamefont {R.-Y.}\ \bibnamefont {Gong}}, \bibinfo {author} {\bibfnamefont {Z.-Y.}\ \bibnamefont {He}}, \bibinfo {author} {\bibfnamefont {C.-H.}\ \bibnamefont {Yu}}, \bibinfo {author} {\bibfnamefont {G.-F.}\ \bibnamefont {Zhang}}, \bibinfo {author} {\bibfnamefont {F.}~\bibnamefont {Nori}},\ and\ \bibinfo {author} {\bibfnamefont {Z.-L.}\ \bibnamefont {Xiang}},\ }\href {https://arxiv.org/abs/2411.19307} {\bibfield  {journal} {\bibinfo  {journal} {arXiv:2411.19307}\ } (\bibinfo {year} {2024})}\BibitemShut {NoStop}%
\bibitem [{\citenamefont {Tzuang}\ \emph {et~al.}(2014)\citenamefont {Tzuang}, \citenamefont {Soltani}, \citenamefont {Lee},\ and\ \citenamefont {Lipson}}]{PS.Tzuang}%
  \BibitemOpen
  \bibfield  {author} {\bibinfo {author} {\bibfnamefont {L.~D.}\ \bibnamefont {Tzuang}}, \bibinfo {author} {\bibfnamefont {M.}~\bibnamefont {Soltani}}, \bibinfo {author} {\bibfnamefont {Y.~H.~D.}\ \bibnamefont {Lee}},\ and\ \bibinfo {author} {\bibfnamefont {M.}~\bibnamefont {Lipson}},\ }\href {https://doi.org/10.1364/OL.39.001799} {\bibfield  {journal} {\bibinfo  {journal} {Opt. Lett.}\ }\textbf {\bibinfo {volume} {39}},\ \bibinfo {pages} {1799} (\bibinfo {year} {2014})}\BibitemShut {NoStop}%
\bibitem [{\citenamefont {Levkovich-Maslyuk}(2016)}]{levkovich2016bethe}%
  \BibitemOpen
  \bibfield  {author} {\bibinfo {author} {\bibfnamefont {F.}~\bibnamefont {Levkovich-Maslyuk}},\ }\href {https://doi.org/10.1088/1751-8113/49/32/323004} {\bibfield  {journal} {\bibinfo  {journal} {J. Phys. A: Math. Theor.}\ }\textbf {\bibinfo {volume} {49}},\ \bibinfo {pages} {323004} (\bibinfo {year} {2016})}\BibitemShut {NoStop}%
\end{thebibliography}%

\end{document}